\documentclass[a4paper,11pt]{article}
\pdfoutput=1
\usepackage{jheppub}
\RequirePackage{luatex85}

\usepackage{amsmath, amssymb, mathrsfs}
\usepackage{mathtools}
\usepackage{slashed}
\usepackage[english]{babel}
\usepackage{graphicx,color,psfrag}

\usepackage{rotating}
\usepackage{array}
\usepackage{slashed,hyperref}
\usepackage{adjustbox}
\usepackage{color}
\usepackage{epstopdf}
\usepackage{overpic}
\usepackage{enumerate}
\usepackage{url}
\usepackage{hyperref}

\def\be{\begin{equation}} 
\def\ee{\end{equation}}
\def\bea{\begin{eqnarray}}
\def\eea{\end{eqnarray}}

\definecolor{orange}{RGB}{255,127,0}
\definecolor{darkgreen}{RGB}{0,127,50}

\title{Reconstructing the spectral shape of a stochastic gravitational wave background with LISA}

\author[a]{Chiara Caprini\,}
\author[b]{, Daniel G.~Figueroa\,}
\author[c]{, Raphael Flauger\,}
\author[d]{, Germano Nardini\footnote{Project coordinator. E-mail: germano.nardini@uis.no}\,}
\author[e,f]{, Marco Peloso\,}
\author[g]{, Mauro Pieroni\footnote{Corresponding author. E-mail: mauro.pieroni@uam.es}\,}
\author[f]{, Angelo Ricciardone\,}
\author[h]{, Gianmassimo Tasinato\,}

\affiliation[a]{Laboratoire Astroparticule et Cosmologie, CNRS UMR 7164, Universit\`e Paris-Diderot, 10 rue Alice Domon et L\'eonie Duquet, 75013 Paris, France.}
\affiliation[b]{Institute of Physics, Laboratory of Particle Physics and Cosmology (LPPC), \'Ecole Polytechnique F\'ed\'erale de Lausanne (EPFL), CH-1015 Lausanne, Switzerland.}
\affiliation[c]{Department of Physics, University of California, San DiegoLa Jolla, CA, 92093.}
\affiliation[d]{Faculty of Science and Technology, University of Stavanger, 4036 Stavanger, Norway.}
\affiliation[e]{Dipartimento di Fisica e Astronomia "G. Galilei",
Universit\`a degli Studi di Padova, via Marzolo 8, I-35131 Padova, Italy.}
\affiliation[f]{INFN, Sezione di Padova, via Marzolo 8, I-35131 Padova, Italy.}
\affiliation[g]{Instituto de F\'isica Te\'orica UAM/CSIC C/ Nicol\'as Cabrera 13-15 Universidad Aut\'onoma de Madrid Cantoblanco, Madrid 28049, Spain.}
\affiliation[h]{Department of Physics, Swansea University, Swansea, SA2 8PP, UK.}

\abstract{We present a set of tools to assess the capabilities of LISA to detect  and reconstruct  the spectral shape and amplitude of a stochastic gravitational wave background (SGWB). We first provide the LISA power-law sensitivity curve and binned power-law sensitivity curves, based on the latest updates on the LISA design. These curves are useful to make a qualitative assessment  of the detection and reconstruction prospects of a SGWB. For a quantitative reconstruction of a SGWB with arbitrary power spectrum shape, we propose a novel data analysis technique: by means of an automatized adaptive procedure, we conveniently split the LISA sensitivity band into frequency bins, and fit the data inside each bin with a power law signal plus a model of the instrumental noise. We apply the procedure to SGWB signals with a variety of representative frequency profiles, and prove that LISA can reconstruct their spectral shape. Our procedure, implemented in the code {\tt SGWBinner}, is suitable for homogeneous and isotropic SGWBs detectable at LISA, and it is also expected to work for other gravitational wave observatories.}

\begin{document}
	
\begin{figure}
\hskip13.cm \href{https://lisa.pages.in2p3.fr/consortium-userguide/wg_cosmo.html}{\includegraphics[width = 0.19 \textwidth]{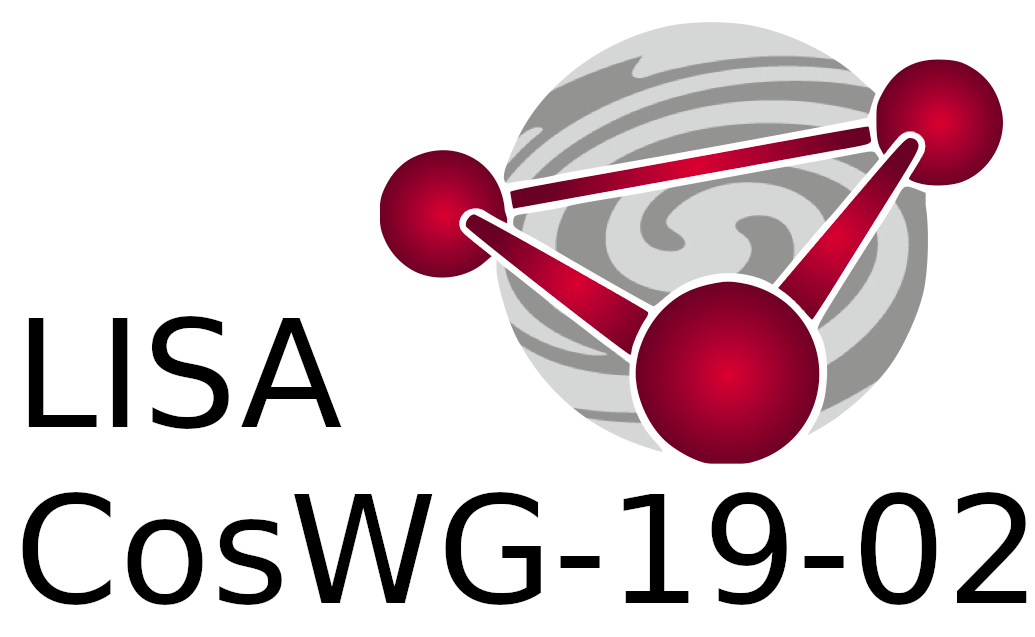}}
\end{figure}

\maketitle
\flushbottom

\section{Introduction}
\label{sec:introduction}

The first gravitational wave (GW) observatory in space, the {\it Laser Interferometer Space Antenna} (LISA)~\cite{Audley:2017drz}, has been  approved by the European Space Agency (ESA) in 2017.  The planned configuration of LISA  has been  fixed to six links, 2.5 million km length arms, and 4 years nominal duration, possibly extensible to 10 years: LISA will be able to probe GWs in the unexplored  milli-Hertz regime. As GWs propagate freely through space, they carry valuable information on the sources that create them, and the geometry of the space-time that they traverse. GWs represent therefore one of the most promising messengers to probe yet unknown aspects of the Universe, which can not be unravelled by other means.

In this paper we discuss the capability of LISA to detect a stochastic GW background (SGWB) and reconstruct the frequency profile of its power spectrum. SGWBs can have both cosmological and astrophysical origin, and each SGWB has typically its own distinctive  profile. The best known example of SGWB of cosmological origin is the irreducible GW background due to quantum vacuum fluctuations during inflation, which spans over a wide range of scales with an almost scale invariant spectrum of small amplitude \cite{Grishchuk:1974ny,Starobinsky:1979ty, Rubakov:1982df,Fabbri:1983us}. Several mechanisms can modulate this tensor spectrum and lead to a non-flat SGWB frequency profile at the scales probed by GW interferometers:  from coupling  the  inflaton  to extra  fields, with a dynamics  characterized by instabilities that enhance the tensor spectrum (see e.g.~refs.~\cite{Cook:2011hg, Senatore:2011sp,Carney:2012pk,Biagetti:2013kwa,Biagetti:2014asa,Goolsby-Cole:2017hod,Sorbo:2011rz, Shiraishi:2013kxa,Anber:2012du,Barnaby:2010vf,Barnaby:2012xt,Ozsoy:2017blg,Maleknejad:2011jw,Dimastrogiovanni:2012ew,Adshead:2013qp,Adshead:2013nka,Obata:2014loa,Maleknejad:2016qjz,Dimastrogiovanni:2016fuu,Agrawal:2017awz,Adshead:2017hnc,Caldwell:2017chz,Agrawal:2018mrg,Espinosa:2018eve}), to models that break space-time symmetries during inflation, leading to a blue spectrum for primordial tensor modes (see e.g.~refs.~\cite{Endlich:2012pz,Bartolo:2015qvr,Ricciardone:2016lym,Ricciardone:2017kre,Domenech:2017kno,Ballesteros:2016gwc,Cannone:2015rra,Lin:2015cqa,Cannone:2014uqa,Akhshik:2014bla,Biagetti:2017viz,Dimastrogiovanni:2018uqy,Fujita:2018ehq}), or scenarios where the  early universe evolution undergoes a brief phase of non-attractor dynamics that amplify the tensor modes \cite{Mylova:2018yap,Ozsoy:2019slf}. These scenarios can lead to a large amplitude of primordial tensor spectrum with a peculiar frequency shape in the frequency band accessible to  experiments like LISA~\cite{Bartolo:2016ami}. Furthermore, post-inflationary, early universe  phenomena can also generate GWs with a large amplitude, e.g.~a kination dominated phase~\cite{Boyle:2007zx,Giovannini:1998bp,Giovannini:1999bh,Figueroa:2018twl,Figueroa:2019paj}, non-perturbative particle production phenomena~\cite{Easther:2006gt,GarciaBellido:2007dg,GarciaBellido:2007af,Dufaux:2007pt,Dufaux:2008dn,Dufaux:2010cf,Enqvist:2012im,Figueroa:2013vif,Figueroa:2014aya,Figueroa:2016ojl,Figueroa:2017vfa,Adshead:2018doq}, oscillon dynamics~\cite{Zhou:2013tsa,Antusch:2016con,Antusch:2017vga,Liu:2017hua,Amin:2018xfe}, strong first order phase transitions~\cite{Witten:1984rs,Kosowsky:1992rz,Kamionkowski:1993fg,Caprini:2007xq,Huber:2008hg,Caprini:2009fx,Caprini:2009yp,Espinosa:2010hh,Hindmarsh:2013xza,Hindmarsh:2015qta,Caprini:2015zlo,Hindmarsh:2017gnf,Cutting:2018tjt,Jinno:2019bxw}, or cosmic defect networks~\cite{Vilenkin:1981bx,Vachaspati:1984gt,Sakellariadou:1990ne,Hindmarsh:1990xi,Damour:2000wa,Damour:2001bk,Damour:2004kw,Figueroa:2012kw,Blanco-Pillado:2017oxo,Ringeval:2017eww,Matsui:2019obe,CosmicStringsLISA}. The resulting GW signal in such cases is given by the superposition of a very large number of uncorrelated and unresolved sources, and hence it is perceived by us as a stochastic background.  For a comprehensive review on SGWB signals of cosmological origin, see ref.~\cite{Caprini:2018mtu}. Most cosmological scenarios are characterized by features in their spectrum that require a more complex parametrization than a single power law, within the LISA frequency band.

On the astrophysical side, various phenomena can also lead to the production of a SGWB in the LISA band. On the one hand, the existence of many compact binaries (black holes, neutron stars, white dwarfs) that will not be resolved as individual sources at LISA, is expected to produce a SGWB contributed by all their GW signals, see e.g.~refs.~\cite{Farmer:2003pa,Meacher:2015iua,Zhu:2011bd,Zhu:2012xw,TheLIGOScientific:2016wyq,Abbott:2017xzg}.   On the other hand, other astrophysical phenomena may contribute as well to produce stochastic backgrounds, from stellar core collapse~\cite{Buonanno:2004tp,Crocker:2017agi,Mueller:2012sv,Ott:2012mr,Kuroda:2013rga}, to r-mode instability of neutron stars \cite{Andersson:1997xt, Friedman:1997uh,Ferrari:1998jf,Zhu:2011pt}, magnetars~\cite{Regimbau:2005ey,Marassi:2010wj}, or superradiant instabilities \cite{Yoshino:2013ofa,Brito:2017wnc,Brito:2017zvb,Cardoso:2018tly,Huang:2018pbu}. See e.g.~refs.~\cite{Schneider:2010ks,Kuroyanagi:2018csn} for general
discussions of astrophysical sources for SGWBs. Under some circumstances, these signals also depend on the events that occurred in the early universe, as is the case of the SGWB sourced by primordial black holes: see refs.~\cite{Carr:2016drx,Garcia-Bellido:2017fdg,Sasaki:2018dmp} for general reviews, and refs.~\cite{Garcia-Bellido:2017aan,Domcke:2017fix,Bartolo:2018evs,Bartolo:2018rku} for examples of studies focussing on
SGWBs detectable at LISA frequencies.

The Universe may be then permeated with a plethora of SGWB signals. Some of the cosmological and astrophysical scenarios  generating SGWBs can occur simultaneously, leading to an overall SGWB with complicated spectral shape. Furthermore, the theoretical uncertainty on the physics of the early universe, and consequently the possibility of the presence of unexpected sources, renders it impossible to predict, from first principles, the exact shape of the cosmological SGWB spectrum over the entire interferometer frequency band. A reasonable expectation is that the SGWB spectrum will likely not follow a single power law in frequency. Given the uncertainty in predicting the expected signal, one cannot apply common techniques of fitting a given model to the data to extract the model parameters, as usually done for e.g.~the Cosmic Microwave Background temperature and polarisation spectra.

From the mere detection of a SGWB signal it will be therefore challenging to identify the source(s) that generated it. Reconstructing the frequency profile, together with other possible features~\cite{Bartolo:2018qqn}, is of paramount importance for this task. One can rely on several templates, but in the lack of a complete catalogue of all possible signal shapes (which seems impossible in view of our limited understanding of the sources), an alternative, unbiased, model-independent reconstruction is compelling. Such a reconstruction should ideally go beyond looking for power-laws or broken power-laws in the whole LISA band, as this neglects key peculiarities of the expected frequency shapes.

So far, most studies adopted the simplified assumption that the signal is well described by a single
power law, characterized by amplitude and slope. Reference \cite{Thrane:2013oya}
introduced the concept of {\it power law sensitivity curve}  (PLS), a graphical representation of the ability of a detector to measure a SGWB with a power law spectrum, for given signal-to-noise ratio and integration time (c.f.~section~\ref{sec:LISA_PlS}). Current SGWB searches focus on power spectra given by a power law of the form $\Omega_{\rm GW}(f)=A(f/f_*)^n$, with $f_*$ a reference frequency~\cite{Lentati:2015qwp,Arzoumanian:2018saf,LIGOScientific:2019vic}. No detection has been made yet, and therefore current analyses provide upper bounds on the amplitude $A$, for different fixed values of the spectral index $n$.  From the recent detection of GW signals from black hole and neutron star binaries~\cite{TheLIGOScientific:2017qsa}, the amplitude of the stochastic GW background from unresolved compact binaries in LIGO/Virgo is estimated as $\Omega_{\rm GW} = 8.9^{+12.6}_{-5.6}\times
10^{-10}$ at $25$ Hz~\cite{LIGOScientific:2019vic}.  It is therefore possible that the SGWB from unresolved binaries might be observed during the next LIGO/Virgo observation runs. 

Reference~\cite{Adams:2013qma} concentrates on LISA -- as we are going to do in this work --  and develops a method to reconstruct both the amplitude and the spectral index of a SGWB power spectrum from the data. The analysis focuses on one single power law signal spanning over the entire LISA frequency band. Two independent reconstructions are considered. The first aims to   fit for the amplitude of the SGWB; the second  fits for both the amplitude $A$ and the spectral index $n$.  Reference \cite{Adams:2013qma}   finds that in the second case the errors on the recovered parameters are inevitably larger, but the SGWB profile can still be reconstructed. 
Accounting for the presence of the foreground due to galactic binaries, ref.~\cite{Adams:2013qma} concludes that a six-link LISA interferometer is able to detect a scale-invariant stochastic background with energy density $\Omega_{\rm GW}= 2 \times 10^{-13}$, with one year of data. In this work, we therefore neglect altogether the presence of the galactic binaries foreground, assuming that this can be subtracted exploiting its yearly modulation~\cite{Adams:2013qma}, and we exclusively focus on the homogeneous and isotropic component. 

The limitations of the power law search are not new. The improvement in the data fits by means of a broken-power-law template has been studied for ground-based detectors~\cite{Kuroyanagi:2018csn, Meyers2018Auth}. Still, the improvement is manifest for idealized signals but not satisfactory for realistic signals. Most of these, indeed, have a more complex form than simple power laws, even within the relatively narrow frequency range of a detector like LISA.

In this work we propose a technique to systematically reconstruct a SGWB signal without  theoretical prejudices on the SGWB frequency profile.  The basic idea is to separate the LISA frequency band into frequency bins, and reconstruct the signal within each bin. For a smooth enough signal, it is a good assumption to approximate it in terms of power laws within small frequency bins. The reconstruction procedure therefore assumes that in every bin the signal is fitted by a single power law (or a constant amplitude), and extracts the best fit values for the amplitude and the spectral index (or only the amplitude). We show that this method can reliably reconstruct signals with non-trivial frequency profiles, taking into account instrumental noise~\footnote{We presented some preliminary results in the proceedings~\cite{Figueroa:2018xtu}. In that document the binning procedure was employed to reconstruct the signal in the case that the noise curve is known.}. We do this by means of an  algorithm that we have implemented in \texttt{SGWBinner}, a code (based on \texttt{Python3}) that automatically performs an appropriate binning of the LISA frequency band, optimized for each given signal profile. 
 
 Specifically, our work is organized as follows. In section~\ref{sec:SGWB_at_LISA} we present the most updated LISA strain sensitivity curve, based on the now established LISA configuration, and construct the PLS of LISA, as given in ref.~\cite{Thrane:2013oya}, with reasonable choices for the SNR threshold and integration time (see \cite{LISA_docs} for all relevant  up to date LISA documents, and in particular \cite{LISA_PLS} for a direct download of the {\it Science Requirements Document}). We account for the fact that, due to measurement breaks needed for the antenna repositioning and other operations, the data taking efficiency of the LISA mission is $\sim 75\%$ of the nominal time, e.g.~out of one year of flight only about 9 months of data will be collected. The 4-year official duration of the mission therefore effectively amounts to three years of data. In section~\ref{sec:SGWB_at_LISA} we also provide a simple, graphical method to predict whether a SGWB with arbitrary spectral shape might be reconstructed by LISA, by dividing the LISA band in several bins, and by computing the PLS curve within each bin. 

 In section \ref{sec_reconstruction} we go  beyond the PLS approach, and develop our binning procedure to reconstruct a SGWB signal with arbitrary spectral shape within the LISA frequency window.  For a given data stream, that  includes signal and  instrumental noise, we provide an algorithm able to determine the best fit for signal and noise in each bin. From this information, our method can then reconstruct the best fit (with associated error bars) for the SGWB spectrum profile in the LISA frequency band. 

 In section \ref{sec_mock} we apply our algorithm, as implemented in the \texttt{SGWBinner} code, to reconstruct several examples of benchmark SGWB signals. The benchmark signals represents illustrative cases leading to a {\it detectable} SGWB in the LISA frequency band. For all benchmark signals, we demonstrate that the agnostic reconstruction of their frequency profile performed by the \texttt{SGWBinner} code, is better (following the AIC, c.f.~section \ref{sec:methodology}) than fitting the data with a single-power-law model. This demonstrates that our algorithm can be a useful tool to reconstruct the amplitude and shape of a SGWB. Consequently, in the case of a stochastic signal detection, our method will be capable of distinguishing SGWBs of different origin. 

 In section~\ref{sec:SNR_thr} we determine SNR$_{\rm thr}$, the minimal SNR to detect a power law SGWB in LISA. We employ our algorithm without splitting the LISA frequency band. With this analysis, we confirm the findings of~ref.\cite{Adams:2013qma}, which obtained the required SNR$_{\rm thr}$ in the case of a flat signal, and we extend their study to the case of a power law signal with non-vanishing slope. In section~\ref{sec_conclusions} we summarize our results, and also mention possible applications and extensions of our work.

\section{Stochastic gravitational wave background detection with LISA}
\label{sec:SGWB_at_LISA}

Once a prediction for a SGWB signal has been formulated, the first concern is to understand whether it is detectable and, in such case, how well its power spectrum can be reconstructed. It is however not easy to answer these questions. For the case of LISA, a robust, precise answer would need to run a complete pipeline on mock data, generated simulating realistic satellites orbits, laser instabilities, plausible glitches and gaps, interruptions of the data stream, the subtraction of all the sources identified individually, and other possible issues. The sophisticated tools necessary to model the data stream in all details will be prepared by the LISA Consortium in the next decade. Here we attempt to provide  a preliminary  answer to the aforementioned questions, estimating the LISA potential to reconstruct the power spectrum of a SGWB signal.

As a first step we focus on a simplified scenario, and we work under the following assumptions:
\begin{itemize}
\item Our data are  the sum of the injected SGWB signal and the (simplified) instrumental noise, i.e.~we consider the ideal case in which the data have been perfectly cleaned from all resolvable sources, glitches,  and any other impurities.
\item LISA data are expected to be acquired in chunks of around 11 days, and we assume that the instrumental operations  between a chunk and the other has no effect on the data. Nevertheless, due to these interruptions, our data are simulated only for 75\% of the duration of the mission, i.e.~3 years out of 4.
\item  The noise and the signal are Gaussian, stationary, and 
uncorrelated in frequency domain. 
\item  The instrumental noise is  described by a model which parameter values are known within about 20\%.
\item  The instrumental noise we use is the one of only one detector channel, the X TDI (Time Delay Interferometry) channel. More realistic data would include the simulation of all X, Y and Z channels, the diagonalisation of the noise matrix to extract the A, E and T channels, and the use of the Sagnac T channel to characterize the noise in the other two (see e.g.~ref.~\cite{Adams:2013qma}). However, under the approximation that the noise and the response function of the A and E channels are the same, we expect that including both channels will increase the SNR by roughly a factor of $\sqrt{2}$ with respect to the single channel case.
\end{itemize} 

\subsection{Model of the LISA sensitivity curve}
\label{sec:LISAsens}

We adopt here the single TDI output noise model of ESA's science requirement document \cite{LISA_docs} (see also ref.~\cite{Cornish:2018dyw} for a derivation). This is based on the results of the LISA Pathfinder mission, which demonstrated that the noise between $\sim\!3\times 10^{-5}$ and $\sim\!10^{-3}\,$Hz can be kept under control~\cite{Armano:2018kix}. The noise at higher frequencies, which depends e.g.~on the laser and optics performances, has been also investigated with a good understanding of the major noise contributions~\cite{LISA_docs}. Based on these facts, the noise of one TDI channel can be  established, within an uncertainty of around 20\%, in the frequency range $3\times 10^{-5}$\,Hz -- $0.5$\,Hz. The simplified noise model we adopt here is based   on the assumption that all arm lengths are constant and equal, and that noises of the same type have the same {\it power spectral density} (PSD). All noise components are then absorbed into two effective functions. The components that dominate the noise at high frequencies are represented by the one-link ``optical metrology system" noise PSD $P_{\rm oms}(f,P)$, whereas the low-frequency components by the single ``mass acceleration" noise PSD $P_{\rm acc}(f,A)$:
\begin{equation}
	\begin{aligned}
	P_{\rm oms}(f, P) &= P^2~ \frac{{\rm pm}^2}{\rm Hz}\left[1 + \left(\frac{2\,\textrm{mHz}}{f} \right)^4  \right] \left(\frac{2 \pi f}{c} \right)^2\; , \\
	P_{\rm acc}(f, A) &= A^2~ \frac{{\rm fm}^2}{{\rm s}^4\,{\rm Hz}}\left[1 + \left(\frac{0.4\,\textrm{mHz}}{f} \right)^2  \right] \left[1 + \left(\frac{f}{8\,\textrm{mHz}} \right)^4  \right] \left(\frac{1}{2 \pi f } \right)^4 \left(\frac{2 \pi f}{c} \right)^2 \; ,
	\end{aligned}
\label{eq:noise-AB}
\end{equation}
with $f$ being the frequency, $c$ the speed of light, and $P=15$, $A=3$ are noise parameters, known to within 20\% \cite{LISA_docs}. The TDI X channel single-sided PSD becomes then
\begin{equation}
 P_{n}^{(X)}(f,P,A) = 16 \sin^2\left(\frac{2 \pi f L}{c}\right) \left\{ P_{\rm oms}(f, P) +\left[3 +\cos \left(\frac{4 \pi f L}{c} \right)\right] P_{\rm acc}(f, A) \right\}\,,
\end{equation}
where $L=2.5\cdot 10^{6}$ km is the arm length, and the $\sin^2(2\pi f\,L/c)$ factor appears as a consequence of the TDI procedure (as shown later, this factor cancels with the response function, c.f.~eq.~\eqref{eq:resp}).  

From the noise PSD of the TDI-X observable, under the above-mentioned assumptions, one can construct the detector strain sensitivity curve for a single TDI channel. This involves the detector polarisation- and sky-averaged response function $\mathcal{R}(f)$, as follows:
\begin{equation}\label{eq:Sn}
     S_{n}(f)=\frac{P_{n}^{(X)}}{\mathcal{R}(f)}\,.
 \end{equation} 
The response of a detector to the GW signal at time $t$ and position ${\bf x}$ is given by the convolution of the detector response with the metric perturbation:
\begin{equation}\label{eq:r}
r({\bf x},t)=\int_{-\infty}^{\infty} d\tau\int d^3y \,R_{ij}({\bf y},\tau)\,h_{ij}({\bf x-y},t-\tau)\,,
\end{equation}
where $R_{ij}$ is the detector response function encoding the time delay measured by the interferometer and it depends on the particular detector design; for LISA see e.g.~ref.~\cite{Romano:2016dpx}.

The metric perturbation can be decomposed as
\begin{equation}\label{eq:h}
    h_{ij}({\bf x},t)=\int_{-\infty}^{\infty} df\int d\Omega_{\hat k}\sum_p h_p(\hat k, f)\,e_{ij}^p \,e^{2\pi i f(t-{\bf x}\cdot \hat k)}\,,
\end{equation}
where $p$ denotes the polarization index, $e_{ij}^p$ the polarization tensors, and $h_p(\hat k, f)$ are the tensor mode functions, whose power spectrum gives the gravitational wave strain PSD $S_{h}(f)$ 
\begin{equation}\label{eq:Sh}
 \langle h_p(\hat k, f)h_q^*(\hat k', f')\rangle =\frac{1}{8\pi}\delta(f-f')\delta(\hat k -\hat k')\delta_{pq}\,S_{h}(f)\,. 
\end{equation}
By introducing the Fourier transform of the response function and the gravitational wave, eq.~\eqref{eq:r} becomes
\begin{equation}
    r(f)=\int d\Omega_{\hat k}\sum_p R_p(\hat k, f) \,h_p(\hat k, f)\,e^{-2\pi i f{\bf x}\cdot \hat k}\,.
\end{equation}
The power spectrum of the detector response in frequency domain $r(f)$ defines the detector response PSD due to gravitational waves:
\begin{equation}\label{eq:Pr}
    \langle r(f)r^*(f')\rangle =\delta(f-f')\, P_{r}(f)\,,~~~~~ P_{r}(f)=\mathcal{R}(f)\,S_h(f)\,.
\end{equation}
The last equation follows from the definition of the polarisation- and sky-averaged response function:
\begin{equation}
    \mathcal{R}(f)=\frac{1}{8\pi}\int d\Omega_{\hat k}\sum_p R_p(\hat k, f) \,R_p^*(\hat k, f)\,.
\end{equation}
A good approximation to the full response function of the LISA instrument is~\cite{int}
\begin{equation}\label{eq:resp}
\mathcal{R}(f) \simeq 16 \sin^2\left(\frac{2 \pi f L}{c}\right)\frac{3}{10} \frac{1}{1+0.6(2\pi f L /c)^2} \left(\frac{2 \pi f L}{c}\right)^2\,.
\end{equation}
The signal-to-noise ratio to a SGWB, accounting for one TDI LISA channel, is given in terms of the detector response PSD and the single TDI output noise PSD (see e.g.~ref.~\cite{Romano:2016dpx} for a derivation in the case of two-detector cross-correlation):
\begin{equation}
{\rm SNR} = \sqrt{T \int_{0}^{\infty} d f \, \left( \frac{P_r (f)}{P_n^{(X)}(f)}  \right)^2 } = \sqrt{T \int_{0}^{\infty} d f \, \left( \frac{S_h (f)}{S_n(f)}  \right)^2 }\,, 
\label{eq:SNR_definition_S}
\end{equation}
where in the second equality we have used \eqref{eq:Pr} and definition \eqref{eq:Sn}. This expression motivates the presence of the response function in the definition of the detector strain sensitivity \eqref{eq:Sn}. 

\begin{figure}\centering
    \includegraphics[width=0.7\textwidth]{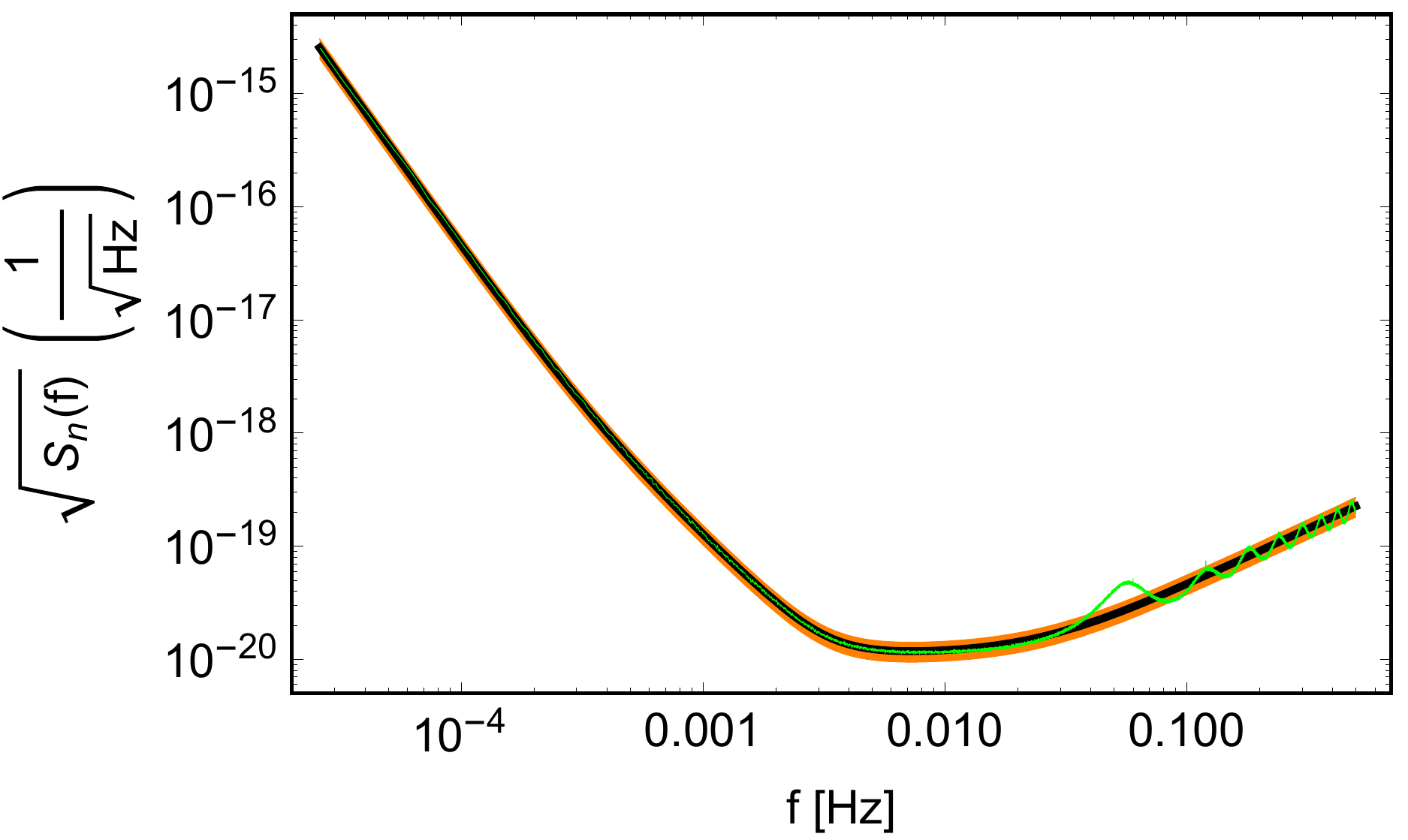}
    \caption{\it LISA strain sensitivity curve for the TDI-X channel: Green, the output from the LISA simulator; Black, the analytical evaluation given in eq.~\eqref{eq:Sn}. The orange band shows the allowed margin on the noise parameters $P=15\pm 20\%$ and $A=3\pm 20\%$.}
    \label{fig:sens_curve}
\end{figure}

The LISA strain sensitivity curve for the TDI X-channel given in eq.~\eqref{eq:Sn} is shown in fig.~\ref{fig:sens_curve}, together with the strain sensitivity output from the LISA simulator. The analytical curve \eqref{eq:Sn} depends on the noise parameters $A$ and $P$ which, as mentioned earlier, have a margin of about 20\%. In fig.~\ref{fig:sens_curve} we also show the strain sensitivity to within this precision, which constitutes the Gaussian prior on $A$ and $P$ that we will use in the following.

Note that the noise model described above plays a key role in our study, since we assume that it perfectly fits the real instrumental noise, and thus we use it to simulate the mock data. If a better modelling of the LISA noise  will be formulated in the future, it can be inserted in the \texttt{SGWBinner} algorithm, which will use it to simulate new data and adopt it in the complete pipeline. The arguments we develop here can therefore be adapted to improved settings of the LISA instrument.

\subsection{The power-law sensitivity curve}
\label{sec:LISA_PlS}

Since in this work we are dealing with a stochastic GW background, it is more common to express the signal in terms of the GW energy density power spectrum, which can be related to the GW strain PSD defined above through (see  e.g.~ref.~\cite{Caprini:2018mtu})
\begin{equation}\label{eq:Omgw}
    \Omega_{\rm GW}(f)=\frac{4\pi^2}{3H_0^2}f^3S_h(f)\,,
\end{equation}
where $H_0=100 \,h \,{\rm km}\,{\rm sec}^{-1}{\rm Mpc}^{-1}$ is the Hubble constant, and we choose $h\simeq 0.67$ \cite{Aghanim:2018eyx}. The GW energy density power spectrum encodes  the relative contribution of the GW to the energy density in the Universe per log-frequency interval:
\begin{equation}
  \Omega_{\rm GW}(f) \equiv \frac{1}{3 H_0^2 M_p^2} \, \frac{\partial \rho_{\rm GW}}{\partial \ln f} \,,
\end{equation}
with $M_p$ the reduced Planck mass. Similarly, one can define the    
 energy density sensitivity  $\Omega_{s}(f)$ through the detector strain sensitivity as~\cite{Maggiore:1900zz}
\begin{equation}\label{eq:Oms}
  \Omega_{s}(f)=\frac{4\pi^2}{3H_0^2}f^3 S_{n}(f) \,.
\end{equation}

The signal to noise ratio to a SGWB is given in terms of the above quantities as (c.f.~eq.~\eqref{eq:SNR_definition_S}):
\begin{equation}
{\rm SNR} = \sqrt{T \int_{f_{\rm min}}^{f_{\rm max}} d f \, \left( \frac{\Omega_{\rm GW} (f)}{\Omega_s(f)}  \right)^2 } \,, 
\label{eq:SNR_definition}
\end{equation}
where $f_{\rm min}$, $f_{\rm max}$ denote, respectively, the minimal and maximal frequencies accessible at the detector.
The SNR increases as the square root of the observation time $T$, and benefits from the broad-band nature of the SGWB, since it contains an integration over frequency. In a seminal paper, ref.~\cite{Thrane:2013oya} introduced a graphical tool to visualise the improvement in sensitivity of a detector to a SGWB, due to the aforementioned properties. This is the PLS, a sensitivity curve constructed in such a way that every power law stochastic signal lying above it has SNR larger than a given threshold ${\rm SNR}_{\rm thr}$. A comparison between a SGWB described by a single power law, and the PLS of a detector, allows one to establish whether or not the signal is detectable with ${\rm SNR}_{\rm thr}$, or larger.

The PLS is constructed as follows. One starts from a given coefficient $\beta$ and a power law signal $\Omega_{\rm GW} (f) = C_\beta f^\beta$. One then finds the value of $C_\beta$ that provides a SNR equal to a given threshold value ${\rm SNR}_{\rm thr}$ across all the LISA band 
\begin{equation}\label{eq:SNRthr}
{\rm SNR}_{\rm thr}=\sqrt{T\int_{f_{\rm min}}^{f_{\rm max}}df \,\,\frac{C_\beta^2 \, f^{2\beta}}{\Omega_{s}^2(f)}}\,.
\end{equation}

This computation is repeated for many values of $\beta$ (concretely, for a dense set of values ranging from a large negative slope to a large positive slope). The PLS is then obtained by associating to each frequency ${f}$ the greatest value of $C_\beta {f}^\beta$.  One guarantees in this way that a power law signal that is above this curve has an SNR value greater than ${\rm SNR}_{\rm thr}$.

The choice of what value of ${\rm SNR}_{\rm thr}$ should be required to make sure that the signal is detectable is far from trivial. Previous LISA analyses investigating the detectability of cosmological signals \cite{Bartolo:2016ami,Caprini:2015zlo} have used the results of ref.~\cite{Adams:2010vc,Adams:2013qma}. By applying a likelihood method and appropriate combinations of the TDI variables, ref.~\cite{Adams:2013qma} found that the old LISA configuration used in that work (six links with 5 million km arms) was able to detect and reconstruct the amplitude of a scale-invariant stochastic background with energy density $\Omega_{\rm GW}= 2 \cdot 10^{-13}$, in one year of data. This result can be converted into a corresponding value of ${\rm SNR}$ that would allow amplitude fitting. To find it, we have compared the constant amplitude $\Omega_{\rm GW}= 2 \cdot 10^{-13}$ to various old-LISA PLS evaluated for several values of ${\rm SNR}_{\rm thr}$ over one year, and found that it corresponds to ${\rm SNR}_{\rm thr}=10$. In ref.~\cite{Bartolo:2016ami,Caprini:2015zlo}, we then chose ${\rm SNR}_{\rm thr}=10$ and $T=1$ year to construct the PLS of all six-links LISA configurations analysed. 

The algorithm to assess whether a SGWB with arbitrary shape is detectable, and its shape reconstructible, that we develop in the remainder of this work, can actually be also used to assess a reasonable value for ${\rm SNR}_{\rm thr}$. As demonstrated in section~\ref{sec:SNR_thr}, despite the differences between our analysis and the one in ref.~\cite{Adams:2010vc,Adams:2013qma} (different LISA configurations, use of one single TDI channel vs. full TDI, integration time, ...), using our code we also find that ${\rm SNR}_{\rm thr}=10$ is a reasonable value. 

\begin{figure}
    \centering  
        \includegraphics[width=0.6\textwidth]{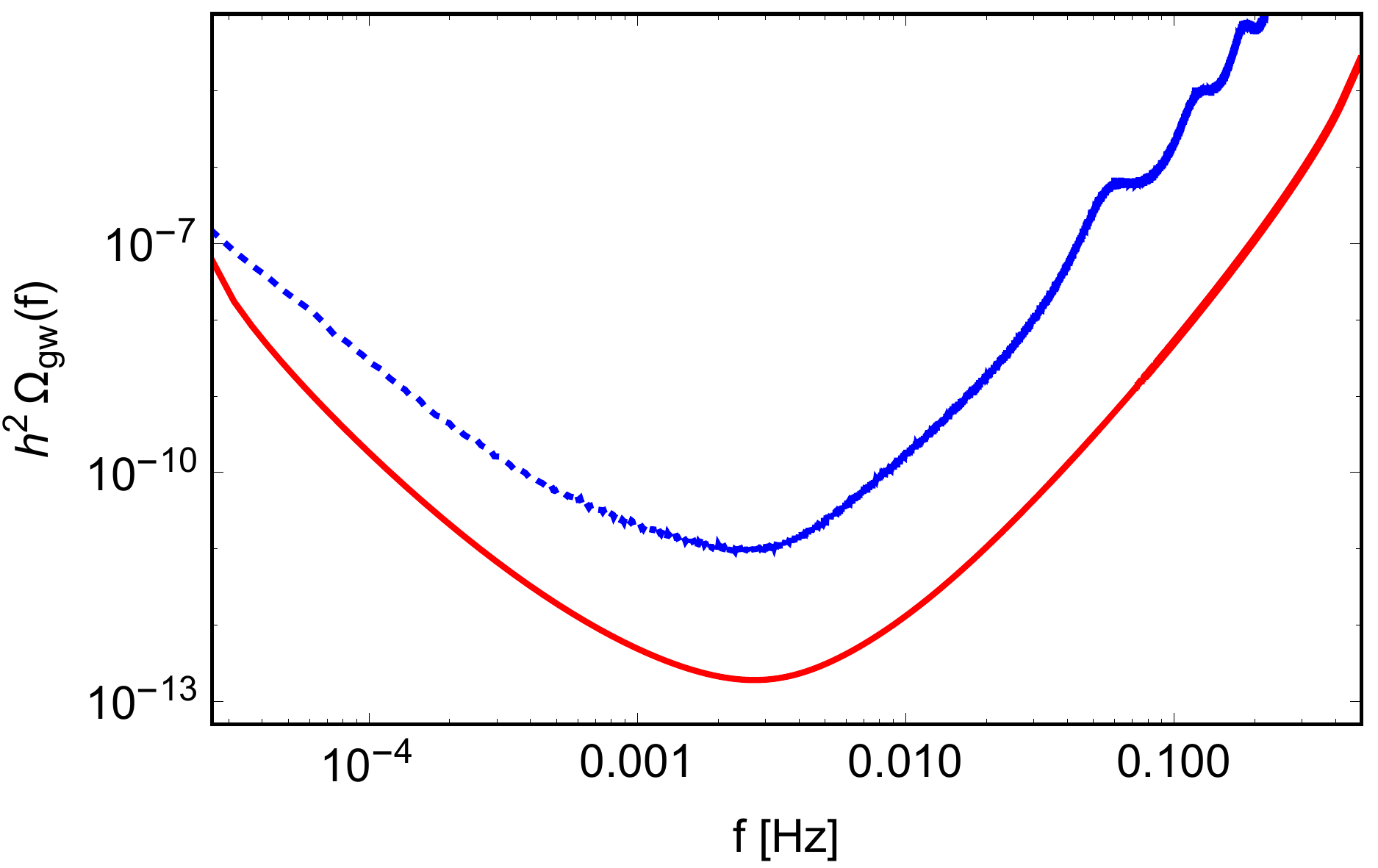}
    \caption{\it Blue, dotted: $h^2\Omega_s(f)$ evaluated with the strain sensitivity $S_n(f)$ output from the LISA simulator (c.f.~fig.~\ref{fig:sens_curve}). Red, solid: the LISA PLS for ${\rm SNR}_{\rm thr}=10$ and $T=3$ year, corresponding to $\sim 75\%$ of the nominal 4-year mission duration, which is the data taking efficiency of LISA.}
    \label{fig:PLS}
\end{figure}

The LISA PLS for ${\rm SNR}_{\rm thr}=10$ and $T=3$ year is shown in fig.~\ref{fig:PLS}, together with the detector energy density sensitivity $\Omega_{s}(f)$ given in eq.~\eqref{eq:Oms}, evaluated with the strain sensitivity $S_n(f)$ output from the LISA simulator (c.f.~fig.~\ref{fig:sens_curve}). We have set $T=3$ years since the data-taking efficiency of LISA will be about 75\%, due to the repositioning of the antenna and other operations, and the nominal mission duration is 4 years (extendable up to 10 years).

\subsection{The binned PLS}
\label{sec:the_multiPlS}

The aim of our work is to develop a technique to accurately reconstruct a SGWB with an arbitrary frequency dependence, typically more complex than a simple or broken power law. In this subsection, we perform a first step in this direction, providing a quick visualisation scheme, that we label the binned PLS, allowing to assess whether a predicted SGWB with arbitrary spectral shape is measurable by LISA (adopting the LISA noise model described in section \ref{sec:LISAsens}). The meaning of the binned PLS is that the shape of any arbitrary SGWB signal can be detected and reconstructed by LISA, if it can be approximated as a series of power laws and it overcomes it.

\begin{figure}
    \centering 
        \includegraphics[width=0.49\textwidth]{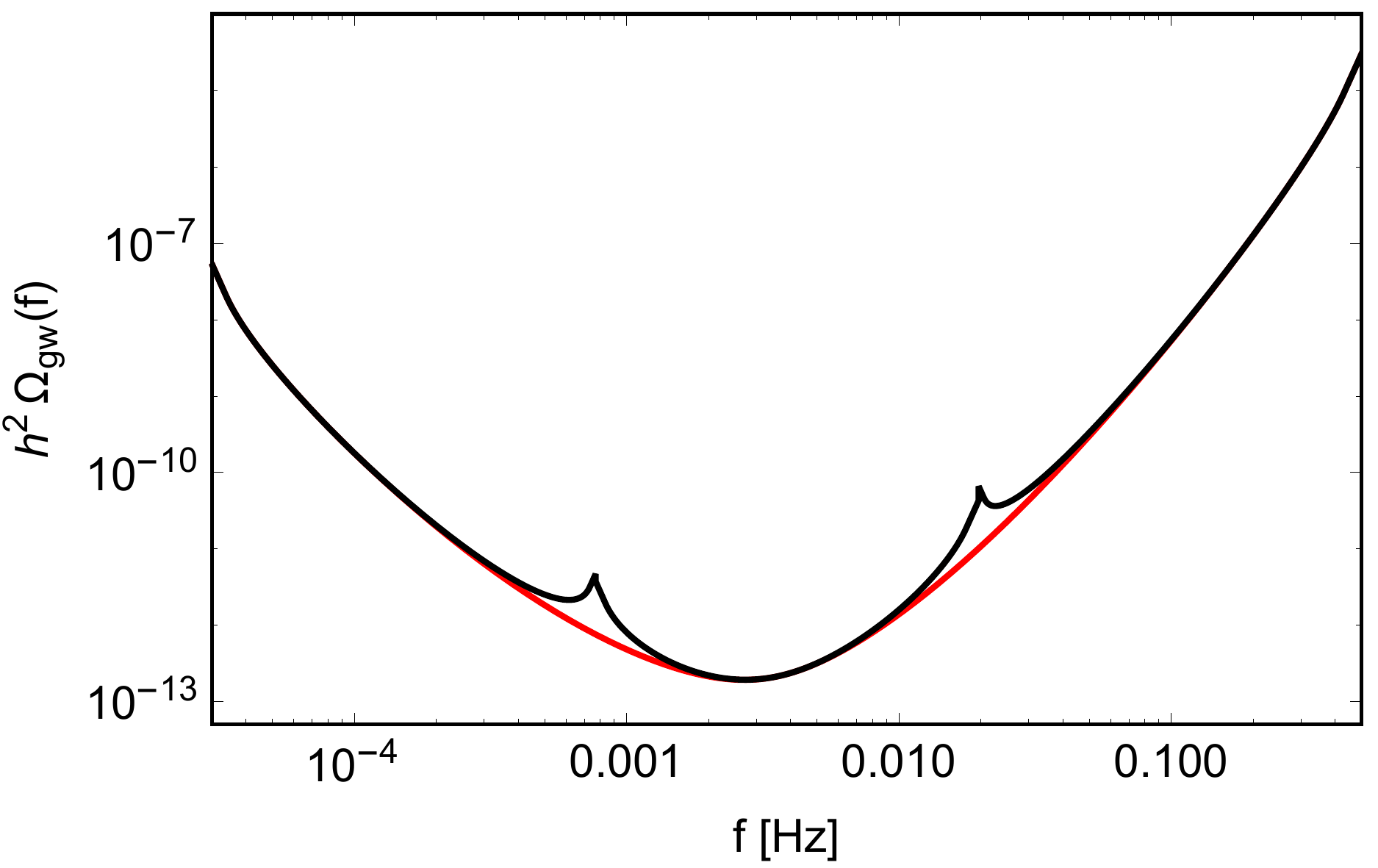}
        \includegraphics[width=0.49\textwidth]{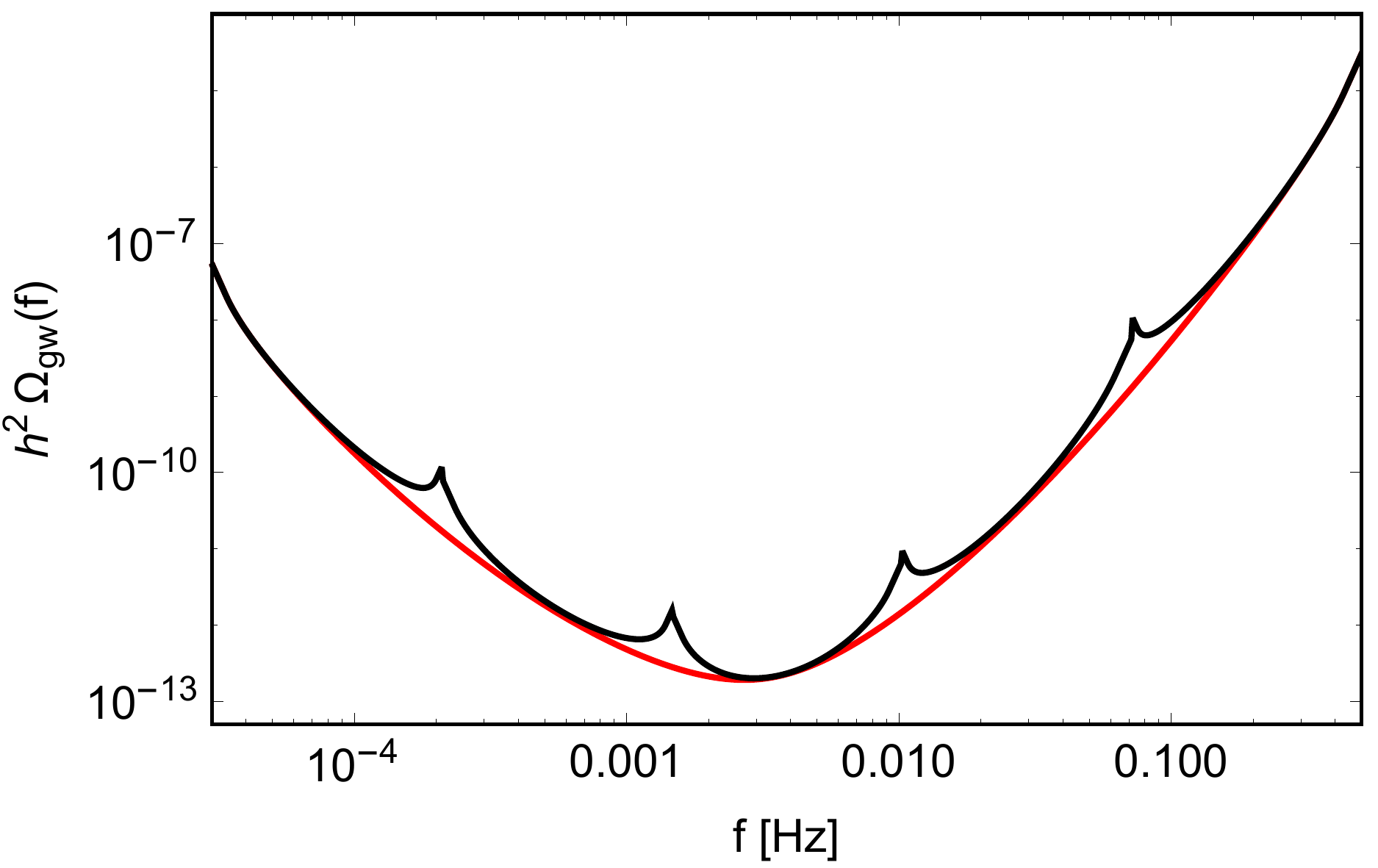}
        \includegraphics[width=0.49\textwidth]{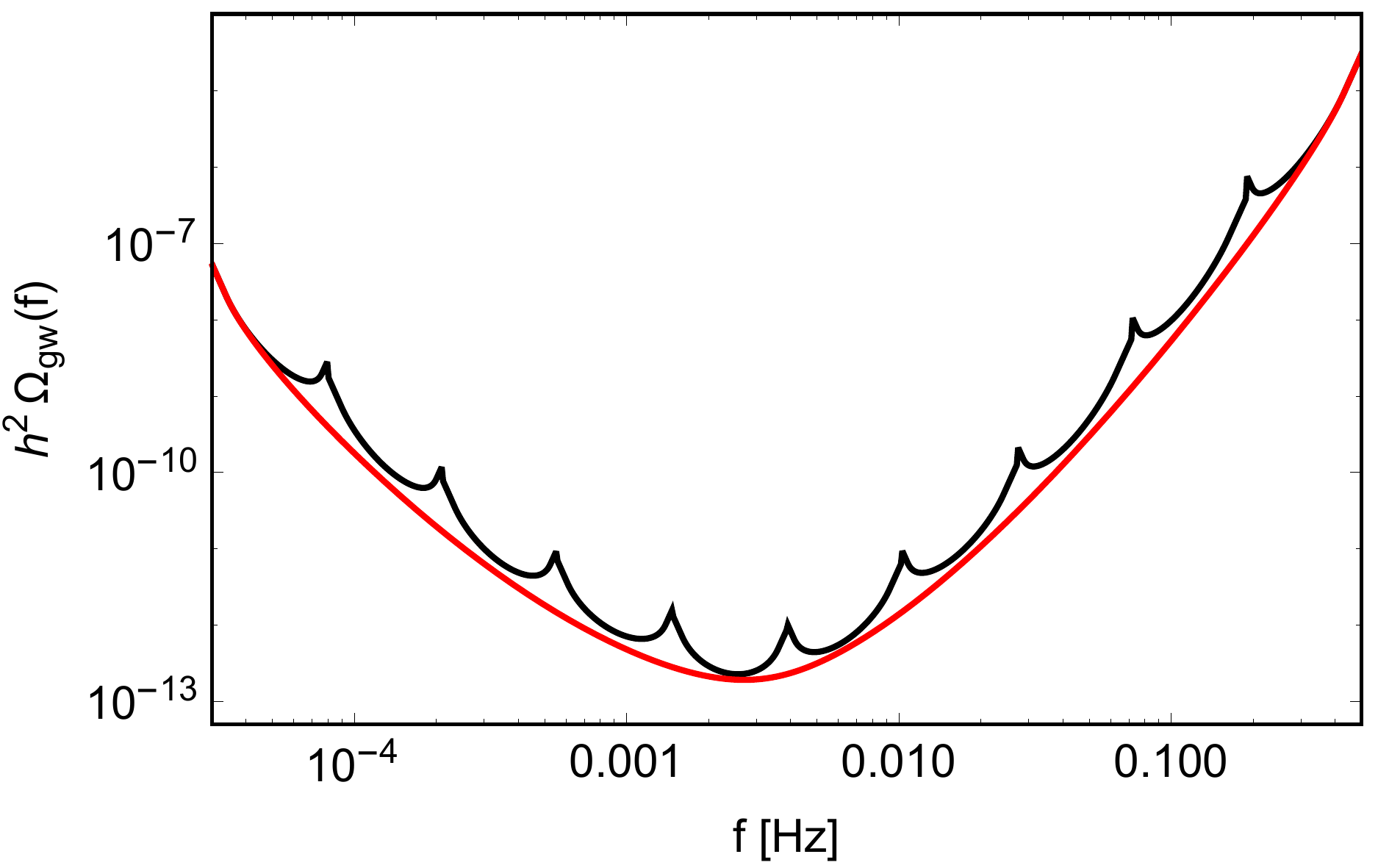}
        \includegraphics[width=0.49\textwidth]{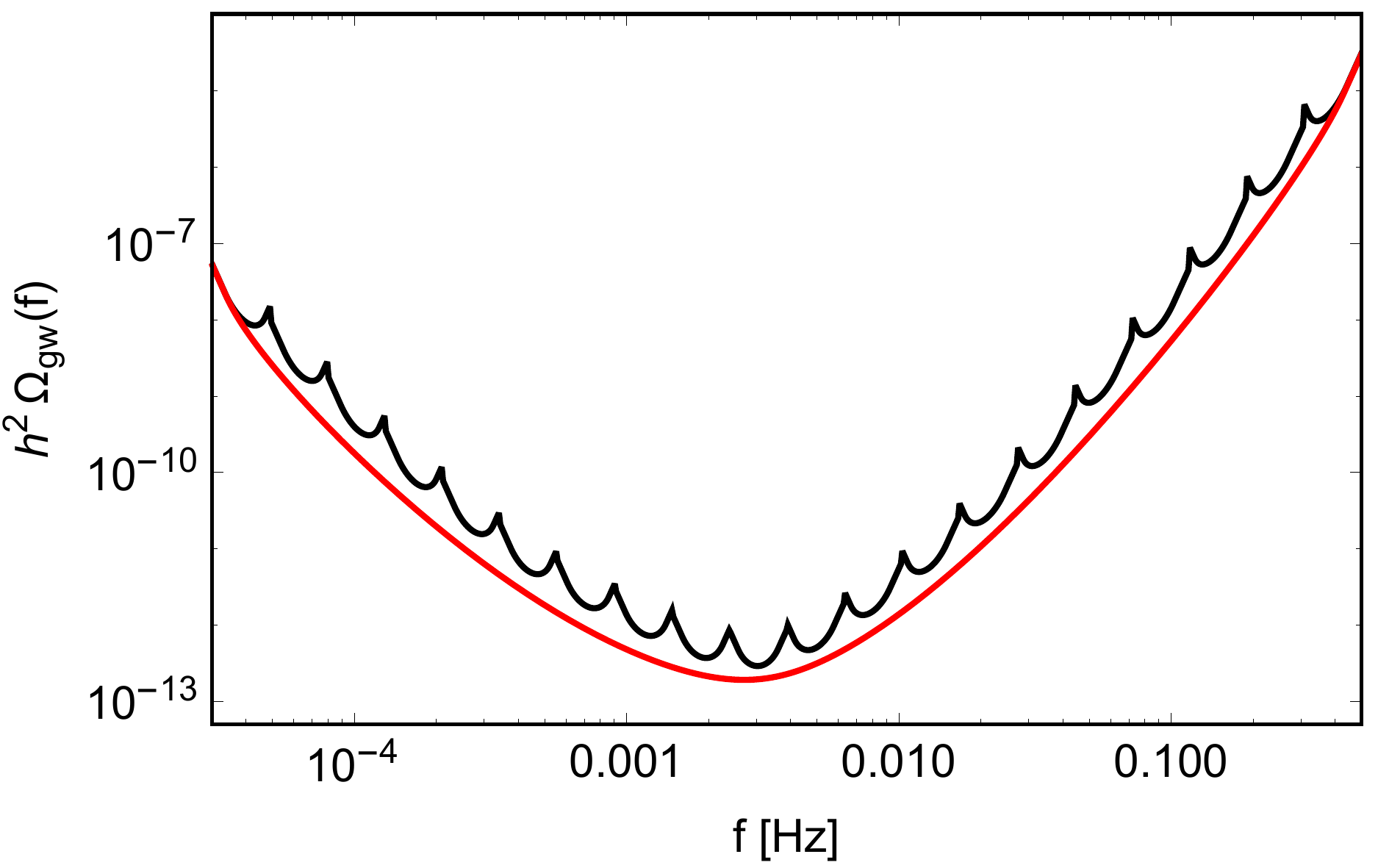}
    \caption{\it Red curve: LISA PLS with ${\rm SNR}_{\rm thr}=10$ and $T=3$ years. Black curve: the binned LISA PLS, from left to right and top to bottom, with $N=3$, $N=5$, $N=10$, $N=20$.}
    \label{fig:PLS_bins}
\end{figure}

For  this aim, we extend the graphical method of section~\ref{sec:LISA_PlS} and construct a sensitivity curve for  an arbitrary  signal  profile by implementing a  binning procedure. We divide the LISA frequency band into smaller intervals, the bins. Assuming that any smooth signal can be well approximated by a single power law within each bin, one can reconstruct the power law amplitude and slope with good accuracy, provided that a sufficient SNR is available in each bin. For each bin we determine the associated PLS curve, and join the binned PLS over the complete LISA band. Figure~\ref{fig:PLS_bins} contains various examples of binned PLS curves for different choices of the number $N$ of frequency bins with the same logarithmic width. The SNR$_{\rm thr}$ is taken to be the same in each bin, SNR$_{\rm thr}=10$ over 3 years. 

In fig.~\ref{fig:PLS_bins} we compare the binned PLS   
with the PLS curve calculated over  the entire LISA frequency band (again with SNR$_{\rm thr}=10$ and $T=3$ years). The only noticeable degradation
in sensitivity occurs at the extrema of each bin. 
This demonstrates that, for a single power law signal spanning over a frequency interval $\Delta f$, the only effectively relevant part of the PLS is the one over $\Delta f$, which in turn justifies the binning procedure. On the other hand, if the number of bins is large (bottom-right panel, $N=20$), the spikes are dense and the binning procedure causes a sizeable effect. For  equal SNR$_{\rm thr}$, the accurate reconstruction of a SGWB with a complicated signal profile, needing many bins, requires a larger average signal amplitude than a simpler signal, requiring less bins.

The binned extension of the graphical method developed in ref.~\cite{Thrane:2013oya} for constructing PLS curves indicates that the analysis of SGWBs with complex spectral shapes is in principle feasible with LISA, provided that sufficient SNR is available. The binned PLS is also a qualitative tool for the model builder, who can simply use it by superimposing the SGWB signal to the curves in fig.~\ref{fig:PLS_bins}. The comparison between the signal and the binned PLS shows at which level the frequency shape of the SGWB signal can be reconstructed in LISA, and whether the signal can be (at least qualitatively) distinguished from those having a different frequency shape. In the remainder of this work we develop a procedure for concretely reconstructing a given SGWB signal with arbitrary frequency profile. 

It is worth commenting on the value of SNR$_{\rm thr}$ we adopt. Our choice of SNR$_{\rm thr}$ is motivated by the fact that, for a power law signal, the data contributing to the SNR are concentrated in a frequency interval that is smaller than the LISA band. The data away from this interval could hence be disregarded and the reconstruction would still be the same. For this reason, at least for a very wide bin, the detection threshold SNR$_{\rm thr}$ in the bin cannot differ from the one used for the single PLS, i.e.~SNR$_{\rm thr}=10$ (see also section \ref{sec:SNR_thr}). However, by reducing the bin width in the binned PLS, at a certain point the bin will stop containing all the relevant data. Still, we keep using SNR$_{\rm thr}=10$ as a criterion for detection. This overlooks the possibility that in a small bin the degeneracy between the signal and the noise might require to increase SNR$_{\rm thr}$. We do not take this refinement into account since, as already stressed, the binned PLS is intended to be a qualitative tool, and small variations of SNR$_{\rm thr}$ would not lead to any appreciable difference in  fig.~\ref{fig:PLS_bins}.

\section{Signal reconstruction beyond simple power laws}
 \label{sec_reconstruction}

In this section we describe a procedure to reconstruct a SGWB signal with frequency dependence more complex than a simple or broken power law (as we demonstrate in the next section, the algorithm of course also works well for single/broken power-law signals).  We divide the discussion in two parts. In the first part,  we describe the methods and the statistical tools at the basis of  our procedure. In the second part, we discuss our algorithm in greater detail. The procedure explained in this section is implemented in a {\tt Python3} code, {\tt SGWBinner}.

\subsection{Methodology} 
\label{sec:methodology}

We aim to reconstruct the frequency dependence of the gravitational wave energy density  $\Omega_{\rm GW}(f)$, for an arbitrary SGWB spectral shape. We divide the entire LISA frequency band in bins, and we  determine the signal frequency profile in each bin. We assume that in each bin the signal is approximated in terms of  a power law. 
 The total signal measured by the instrument is the uncorrelated sum of the gravitational wave signal 
 $\Omega_{\rm GW}$ and noise $\Omega_s$, 
 \begin{equation} 
h^2\,\Omega_{\rm tot} \,=\,h^2\,\Omega_{\rm GW}  +h^2\,\Omega_s \,. 
\label{omega-tot}
\end{equation}
The noise model determining $\Omega_s$ has been presented in section \ref{sec:LISAsens} (see also section~\ref{sec:LISA_PlS}). We now briefly discuss the theoretical models adopted to reconstruct the signal $\Omega_{\rm GW}$. 

 \smallskip
 \noindent
{\bf Signal model:} We  consider a piece-wise signal characterized by two possible parameterizations within the bins:
   \begin{enumerate}
       \item  
   The first parametrization assumes that the signal is 
 fitted  in terms of  
an amplitude and a slope within each bin, denoted by the index $i$ (with $\vec s_i = (\alpha_i,p_i)$ indicating amplitude and slope)
\begin{equation}
h^2\,\Omega_{\rm GW} \left( f ,\, \vec{s}_i \right) =  \;  10^{\alpha_i}  \, \left( \frac{f}{\sqrt{f_{{\rm min},i} \, f_{{\rm max},i}}} \right)^{p_i}  
 \; \theta \left( f - f_{{\rm min},i} \right) \,  \theta \left( f_{{\rm max},i} - f \right) \,, 
\label{2-param}
\end{equation} 
where  $\theta$ is the Heaviside step function,
 $f_{\rm min, \,max}$ are the bin extremal frequencies,  and
  we use the geometrical mean of the endpoints of each bin to determine the pivot scale for the slope in each bin. 
  
 \item    
The second parametrization  assumes  a constant amplitude and zero slope ($p_i\,=\,0$) in each bin: 
\begin{equation}
h^2\,\Omega_{\rm GW} \left( f ,\, \vec{s}_i \right) = 10^{\alpha_i}  \; \theta \left( f - f_{{\rm min},i} \right) \,  \theta \left( f_{{\rm max}, i} - f \right) \,.
\label{1-param}
\end{equation}
\end{enumerate} 
We denote these two parameterizations  as the ``2-parameter'' ``1-parameter" fits, respectively.

\smallskip

Our method of reconstruction will be based on a procedure of  maximization  of the likelihood function. We generate
 a random realization for noise and signal; we then divide the entire  LISA frequency band into a number of equally log-spaced intervals, the bins, and we  coarse grain over our data:  the  procedure we follow is discussed in more detail in the next subsection. 
 The {\it posterior probability}  describing the distribution of our theoretical parameters is 
 given by the expression (up to a normalization factor independent of the signal and noise parameters, and hence irrelevant for
the reconstruction procedure)
\begin{equation}
\mathcal{L}_{\rm tot} \left( \vec{s},\,\vec{n} \right) 
 \,\propto  \, p_N \left( \vec{n} \right) \,p_S \left( \vec{s} \right) \, \times\,  \mathcal{L} \left( \vec{s},\,\vec{n} \right) \; ,
\end{equation}
 where $p_N$ and $p_S$ are, respectively, our priors on the noise and on the signal parameters,  while  $ \mathcal{L}$  is the likelihood (the explicit form of which is given in  subsection  \ref{sec:implementation}). The signal parameters are collectively indicated as the vector  $\vec s$ containing the set of amplitudes and slopes for all bins (or only amplitude in the case of the second parametrization), while the noise parameters are indicated as the vector $\vec{n} = \left\{ A ,\, P \right\}$ (c.f.~section \ref{sec:LISAsens}).  We assume a flat prior for the signal, while the prior for the noise will be discussed in the next subsection. From the 
 total likelihood
   we obtain the best-fit values for the signal and noise parameters within each bin, and the confidence level regions around these values. The contour lines can be obtained by the variation of  the exact likelihood  
   or (under the simplifying assumption of a Gaussian $\mathcal L$) via the Fisher information matrix: 
\begin{equation}
C_{ij}^{-1} \equiv -  \, \partial_i \partial_j \ln 
\mathcal{L}_{\rm tot} 
\Big\vert_{best \; fit} \;, 
\end{equation}
where the indexes $i$ and $j$ run over all the (noise and signal) parameters. 
In this way, we are able to reconstructthe best fit for any gravitational wave signal profile  over the entire LISA frequency band, which is divided into small bins where the signal can be approximated as a power law. 
 
\smallskip
 
\noindent
{\bf Merging nearby  bins:} 
As a last important step of our procedure,  we implement a method for testing whether the signal reconstruction is improved by merging nearby bins, and by determining the best fit in larger intervals containing more points. This might be the case if the complete signal is well
 described by  power laws over a large frequency interval.
In such situation, dividing the interval in  many small bins would unnecessarily introduce too many fitting parameters. We employ the 
AIC \cite{Aka74} to determine whether it is convenient to merge two nearby bins. 
For definiteness, let us compare an analysis referred to as {\it A}, performed with $N+1$ bins, against an analysis referred to as {\it B}, in which two nearby bins have been merged (therefore, analysis {\it B} consists of $N$ bins).  For both cases, we compute the quantity
\begin{equation}
{\rm AIC} \equiv \chi_{\rm best \; fit}^2 + 2 \, k \;, 
\label{AIC}
\end{equation}  
where the chi squared is related to the likelihood by $\chi^2 = - 2 \, \ln \, {\cal L}  $, evaluated in the best-fit, while $k$ is the number of parameters used in the fit.

We can interpret  eq. (\ref{AIC}) as the sum of two ``penalty'' factors. Namely an analysis is penalized if it has a larger $\chi^2$ (which is equivalent to a smaller likelihood) or if it employs too many parameters. Therefore, we choose the analysis (between {\it A} and {\it B}) with the lower value of AIC. The code attempts to merge nearby bins in an iterative way, until this does no longer decrease the AIC value. 

\subsection{Implementation} 
\label{sec:implementation}

In this subsection we discuss
 in more detail how the algorithm described in  subsection \ref{sec:methodology}, is implemented in our code {\tt SGWBinner}. For concreteness, we assume that LISA  provides continuous measurements (without the need of any mechanical adjustment such as repositioning of the antenna needed to send the data to Earth) for a period of approximately 11 days: see 
 \cite{Audley:2017drz}. Hence a LISA's $4$ year mission  will produce $94$ such data chunks, corresponding to approximately $75\%$ observing efficiency. It may be possible to combine these chunks into longer data streams;
 conservatively, we analyze the chunks separately and combine their reconstructed power spectra. 

 \vspace*{0.3cm}
\noindent
{\bf Simulation of the data stream.} We start by Fourier transforming the data stream. We  consider frequencies ranging from a minimum frequency of $f_{\rm min} = 3 \cdot 10^{-5} \, {\rm Hz}$ to a maximum frequency  $f_{\rm max} = 0.5 \, {\rm Hz}$ with a frequency spacing of $\Delta f \simeq 10^{-6} \, {\rm Hz}$ set by the length of the time stream. Since the time stream is real, we shall  work from now on with positive frequencies only,  and write the time data stream as
\begin{equation}
d(t)=\sum\limits_{f= f_{\rm min} }^{f_{\rm max}} \left[ d(f)e^{-2\pi i f t}+d^*(f)e^{2\pi i f t} \right]\,.
\end{equation}
We assume that the stochastic gravitational wave background and the noise are stationary,  so that $\langle d(t)d(t')\rangle=f(t-t')$, and have zero mean. The ensemble averages of the Fourier coefficients  satisfy
\begin{equation}
\langle d(f)d(f')\rangle=0\qquad\text{and}\qquad \langle d(f)d^*(f')\rangle=D(f)\delta_{ff'}\,.
\end{equation}
The real and imaginary parts of $d(f)$ are independent random variables with variance $D(f)/2$. The same logic separately applies to the signal and noise. We  further make the hypothesis  that the signal and the noise are Gaussian (so that the power spectra completely characterize their statistical properties). To generate a realization of a simulated signal, at each value of frequency the code generates the quantity
\begin{eqnarray} 
S_i &=& \left\vert \frac{G_{i1} \left( 0 ,\, \sqrt{h^2 \Omega_{\rm GW} \left( f_i \right)} \right) + i \, G_{i2} \left( 0 ,\, \sqrt{h^2 \Omega_{\rm GW} \left( f_i \right)} \right) }{\sqrt{2}} \right\vert^2 \;\;,\;\; \nonumber\\ 
N_i &=& \left\vert \frac{G_{i3} \left( 0 ,\,  \sqrt{h^2 \Omega_{\rm s} \left( f_i \right)} \right) + i \, G_{i4} \left( 0 ,\, \sqrt{h^2 \Omega_{\rm s} \left( f_i \right)} \right) }{\sqrt{2}} \right\vert^2 \,.
\label{N-S-simulated}
\end{eqnarray} 
In this expression $G_{i1} \left( M ,\, \sigma \right) ,\; \dots ,\; G_{i4} \left( M ,\, \sigma \right)$ are $4$ real numbers randomly drawn from a Gaussian distribution of average $M$ and variance $\sigma$, representing the real and imaginary parts of the Fourier coefficients of signal and noise. 

The values of the signal and noise powers are then  added   to form  the data 
 (under the assumption of noise uncorrelated with the signal)
\begin{equation}
D_i = S_i + N_i  \,,
\end{equation}
which corresponds to the relation (\ref{omega-tot}). For each frequency $f_i$, the code produces $94$ values $\left\{ D_{i1} ,\, \dots ,\, D_{i94} \right\}$, and it then computes their average ${\bar D}_i$ and standard deviation $\sigma_i$. The standard deviation $\sigma_i$ is employed as an estimate of the error associated with the measurement at the frequency $f_i$. The
 likelihood function that we adopt is
\begin{equation}
\mathcal{L}(\vec{s},\vec{n}) \propto \exp\left( - N_{\rm chunks} \sum_i  \frac{1}{2} \left[ \frac{{\bar D}_i - h^2 \Omega_{\rm GW} \left( f_i,\vec{s} \right) - h^2 \Omega_s \left( f_i, \vec{n} \right) }{ \sigma_i}\right]^2 \;   \right) \; , 
\label{likelihood} 
\end{equation}
(up to a proportionality factor independent of $\vec{s}$ and $\vec{n}$). The factor $ N_{\rm chunks} = 94$ accounts for the fact that each frequency is measured $ N_{\rm chunks} $ times. 

 \vspace*{0.3cm}
 \noindent
{\bf Coarse graining the simulated data.} Given the linear spacing $\Delta f = 10^{-6}$ Hz, the code has a large number of data at the largest frequencies. This considerably increases the computational  time.   For generic signals, we do not expect to need a resolution of ${\rm O } \left( 10^{-6} \right)$ Hz at frequencies much greater than this value. Therefore, the code coarse grains the simulated data, with a coarse graining that increases with increasing frequency. Specifically, the code keeps all the original values from the minimum frequency $f_{\rm min} = 3 \cdot 10^{-5} $Hz to the frequency $f = 10^{-3} $ Hz (this corresponds to the first $971$ frequencies). The code then splits the remaining frequency range (from $f = 10^{-3} $ Hz to the maximum frequency $f_{\rm max} = 0.5 \, {\rm Hz}$), in $1000$ intervals of equal log-spacing~\footnote{The code offers also the alternative option to divide the available frequency  range  in intervals with roughly the same SNR.  In this case,  
in order to agree with the criteria discussed in section \ref{sec:the_multiPlS} (see also section \ref{sec:SNR_thr}) we fix to 10 the minimal SNR$_{\rm thr}$  allowed for each bin.}. In each interval the code obtains a single point, determined by the weighted average of the  frequencies $f_i$ and the simulated value ${\bar D}_i$ of all the points contained  in that interval (where each point is weighted by ${1}/{\sigma_i^2}$). The final, coarse grained  points are used in the likelihood (\ref{likelihood}); for each final point we use the error $\sigma = \left( \sum_i {1}/{\sigma_i^2} \right)^{-1/2}$.

\vspace*{0.3cm}
\noindent
    {\bf Characterization of the noise.}  The next step is to improve the characterization of the noise. To do so, we divide the full frequency range from $f_{\rm min}$ to $f_{\rm max}$ into $5$ intervals of equal log-spacing. We will refer to them as the first, second, third, fourth and fifth intervals, or as the outermost left (first), central (second, third and fourth) and outermost right intervals. We use the outermost left and the outermost right intervals to obtain a prior on the noise, that we later use when we analyse the data in the range of frequencies within the three internal intervals. The reason for choosing these external intervals is that the instrumental noise increases
 in these extremal regions, and we expect the noise to dominate over a generic signal there. With this choice, the outer  $2/5$ths (in log spacing) of the LISA frequency range (where the noise is largest) is `sacrificed' to characterize the noise parameters, while the rest is used to reconstruct the signal. This is  an arbitrary choice, that can be revised for signals that are peaked towards the boundaries of the LISA range.

In section \ref{sec:LISAsens} we explained that the noise is characterized by two parameters, $P$ and $A$, introduced in eqs.~\eqref{eq:noise-AB}, which are assumed to be constant across the LISA band. Specifically,  
the acceleration noise, proportional to $A$, dominates the noise at small frequencies (therefore, in the first interval), while the optical metrology, proportional to $P$, dominates it at high frequencies (therefore, in the fifth interval). The code fits jointly the data in the first and fifth intervals, through a $6$ parameter fit (namely, the parameters $A$ and $P$, common to both intervals, the amplitude and slope in the first interval, and the amplitude and slope in the fifth interval), using the priors  
\begin{equation}
p_N(\vec{n}) = \exp \left\{ -\sum_i \left( \frac{ n_i -\bar{n}_i }{2 \sigma_i} \right)^2 \right\} \; ,\;\; n_1 = A \;,\;\; n_2 = P \;,\;\; \bar{n}_1 = 3 \;,\;\;  \bar{n}_2 = 15 \;,\;\; \sigma_i = 0.2\times  \bar{n}_i \;, 
\end{equation} 
for the noise parameters, motivated by the values entering in eqs.\eqref{eq:noise-AB} for the noise 
functions.

\vspace*{0.3cm}
\noindent
{\bf Reconstructing   the signal and noise parameters.}  The outermost frequency intervals are no longer employed to analyze the data, to avoid using the same data twice. The frequency range covered by the second, third, and fourth interval is then divided in a number $N$ of bins of equal log-spacing. The value of $N$ can  be chosen by the user through an input parameter. 
 For the analysis in this work we 
 tried different values of $N$ between $5$ and $30$. This is a compromise between precision in reconstructing the signal and time needed for the analysis. We verified that, in most cases, the precise value of $N$ is unimportant, thanks to the procedure of bin merging described in subsection \ref{sec:methodology}.  The code then analyses the data in each bin, using either a three parameter (amplitude of the GW signal, noise parameter $A$, noise parameter $P$) or a four parameter fit (amplitude and slope of the GW signal, noise parameter $A$, noise parameter $P$) for each bin, depending on the choice of the user. The best-fit values of $A$ and $P$ obtained in the previous step (using the first and fifth interval) are employed as the initial guess in the minimization of the likelihood within each bin. Moreover, the likelihood obtained in the previous step, with the values of the signal parameters fixed at their best-fit value, is used as a prior for $A$ and $P$ within each bin.   
Once the best fit values have been obtained for all the starting $N$ bins, the code tries (in a recursive manner) to merge neighbouring bins, using the Akaike Information Criterion (AIC), as described at the end of the previous subsection. 

Once the final number of bins is selected, and the best fit value for the parameters are obtained in each bin,  the code computes the posterior in the space of signal parameters. We recall that, depending on the user's choice, this consists of a single parameter (the amplitude) or of two parameters (the amplitude and the slope) per bin. This is done, within each bin, by marginalizing the total posterior over the values of the noise parameters in the specific bin. The marginalization is done analytically, by assuming that the posterior has a Gaussian profile in the directions of the noise parameters (that we reconstruct using the best fit values and the  Fisher covariance matrix). The marginalized posterior is then used by the code 
to generate contour plots (one per bin) in the space of parameters. 
The contour lines around the best-fit values $\vec{\theta}_{\rm best} $ are obtained by the variation of the chi square function (obtained from the logarithm of the marginalized posterior). As standard, the variations $\Delta \chi^2 = 1 ,\, 4$ and $\Delta \chi^2 =  2.30, \ 6.18 $ determine the $1 \sigma$ and $2 \sigma$ contour levels, respectively for the one-parameter (only amplitude) and for the two-parameters (amplitude and slope) cases.    

\vspace*{0.3cm}
\noindent
{\bf Visualization of the reconstructed signal}. We now discuss how the {\tt SGWBinner} code visualizes (bin by bin) the reconstructed signal in the $\left\{ f ,\, h^2 \, \Omega_{\rm GW} \right\}$ plane, starting from the $1$ and $2 \sigma $ confidence level regions for the signal parameters. For the 1-parameter fit (amplitude only), it is immediate to draw a line corresponding to the best fit amplitude in each bin, together with the $1\sigma$ and the $2\sigma$ bands around this amplitude. For the 2-parameter fit (amplitude and slope), consider the points along the $1$ and  $2 \sigma $ contour lines. Each point is specified by one value of
  the amplitude and one value of the slope, and so it is associated to a power law power spectrum (within its bin) given by eq.~\eqref{2-param}. The set of all these power law power spectra (one per point in the contour) covers a region in the $\left\{ f ,\, h^2 \, \Omega_{\rm GW} \right\}$ plane, that surrounds the line that corresponds to the best-fit power spectrum (the one specified by the values $\alpha$ and $p$ that minimize the $\chi^2$). The region covered by these lines is the $1\sigma$ and $2\sigma$ region for that bin. 

\begin{figure}[t!]
\begin{center}
\includegraphics[width=0.75\textwidth]{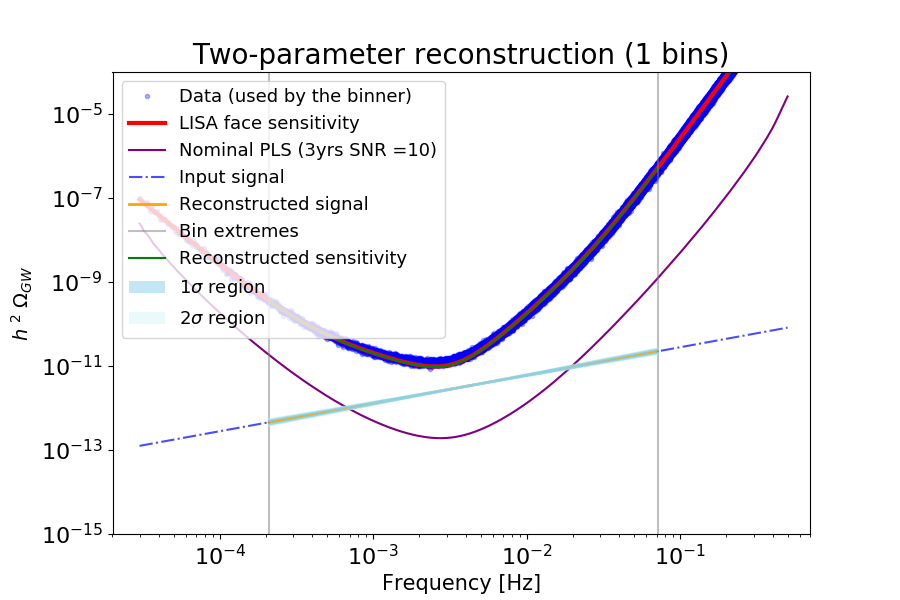}
\end{center}
\caption{\it Simulated data, sensitivity curves, input signal (not visible as it is covered by the error band of the reconstructed signal), and reconstructed signal and sensitivity
 by means of the \texttt{SGWBinner} code. See 
the main text for a detailed explanation.}
\label{fig:example1}
\end{figure}

\begin{figure}[t!]
\begin{center}
\includegraphics[width=0.75\textwidth]{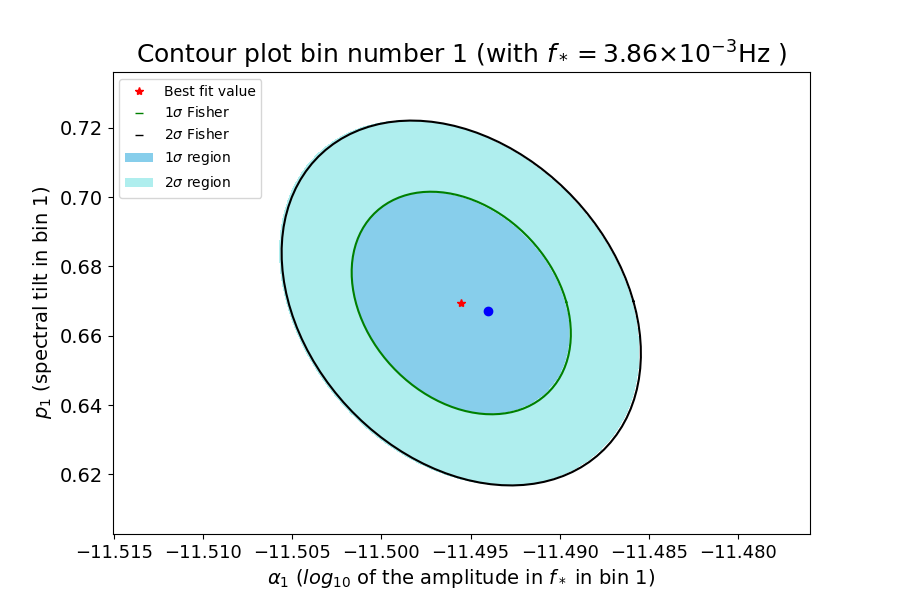}
\end{center}
\caption{\it 
Best fit and $1$ and $2 \sigma$ contour lines for the amplitude and slope of the reconstructed signal in the central bin visible in fig.~\ref{fig:example1}.  The blue mark shows the input signal parameters.}
\label{fig:example2}
\end{figure}

The result of this procedure is shown in fig.~\ref{fig:example1},  \ref{fig:example2}.  In fig.~\ref{fig:example1}, the simulated data appear as a blue thick band (due to their large number, these points cannot be distinguished individually in the figure). The LISA sensitivity curve in energy density $h^2\Omega_s$ (calculated with the analytical noise model of section \ref{sec:LISAsens} with $P=15$ and $A=3$) is shown with a red curve. The input signal is a power law with slope $2/3$ and with amplitude $1.3 \cdot 10^{-12}$ at the frequency $ 0.001 \, {\rm Hz}$. The vertical lines placed at $f_L = 0.00021$ Hz and at $f_R=0.07125$ Hz  are the boundaries between the two external regions used to obtain priors on the noise parameters, and the three internal regions used to analyze the signal.   (These frequencies correspond to $1/5$th and $4/5$th of the total interval in log units, within the numerical precision). 
 
  We then divided the range included in the internal regions into $N=5$ initial bins. The merging procedure based on the AIC then found convenient to merge these bins into a single bin, spanning the full range between $f_L$ and $f_R$. The best fit signal reconstructed by the code is the power law shown as a solid yellow line. The light blue band around this line contains the $1$ and $2\sigma$ region, determined by the procedure explained in the previous paragraph. Finally, the figure also shows with a green line the best-fit LISA sensitivity reconstructed by the code in the internal bin.

Let us conclude with a description of the contour plots in fig.~\ref{fig:example2}. The red star marks the best fit value in the internal bin.
The green and black ellipses are, respectively, the $1$ and $2 \sigma$ contour lines around this best fit value, determined with the Fisher matrix. The light blue regions are instead the  $1$ and $2 \sigma$ areas determined from the exact  likelihood ${\cal L}_{\rm tot}$. The very good agreement between these regions and the contour lines shows that the shape of the exact likelihood is well approximated by a Gaussian, close to the best-fit value. We recall that the amplitude of the reconstructed signal $\alpha_i$ is given at a frequency which is the geometrical mean of the corresponding interval. In this case, this corresponds to the pivot frequency $f=0.003868 \, {\rm Hz}$.  The log$_{10}$ values of the amplitude $\alpha_i$ at this frequency and of the slope of the input signal $p_i$ are $\left\{  -11.4944 ,\, 0.667 \right\}$. 
We can see from fig.~\ref{fig:example2} that these values lie inside the best fit $1 \sigma$ area.

\section{Reconstructing mock signals}
\label{sec_mock}

As anticipated in the introduction, theoretical predictions on  SGWBs of  cosmological or astrophysical origins      encompass a rich variety of frequency profiles, with spectral shapes ranging from  a simple power law to peaked signals, passing through monotonic signals with smoothly growing or decreasing slopes, formed by two or various broken power laws, and so on. 
In this section we apply the {\tt SGWBinner} code to several mock data sets constructed using the LISA noise model and several SGWB templates, which have high enough amplitude to be detectable.

\subsection{Benchmark signal shapes}

On the astrophysical side, we expect SGWBs within the frequency range of LISA, from the superposition of the unresolved GW emission from 
compact binaries: galactic binaries, extra-galactic stellar origin black hole binaries (BHB), extra-galactic neutron star binaries (NSB), and so on. For example, the LIGO-Virgo background due to NSBs and BHBs, in the frequency band most sensitive to SGWBs (around 25 Hz), is currently estimated to $\Omega_{\rm GW}(f = 25 ~{\rm Hz})=8.9^{+12.6}_{-5.6}\times
10^{-10}$~\cite{LIGOScientific:2019vic}. 
Concerning the SGWB from galactic binaries in the LISA band, here we make the simplifying assumption that it can be subtracted from the data stream, with techniques exploiting its yearly modulation, as done e.g.~in \cite{Adams:2013qma}. Cosmological sources can provide SGWBs with power law  spectra characterised by  different slopes than the astrophysical ones: e.g.~in the presence of a Kination dominated phase~\cite{Boyle:2007zx,Giovannini:1998bp,Giovannini:1999bh,Figueroa:2018twl,Figueroa:2019paj} the spectrum scales as $\propto f^\alpha$, with $0.5 \lesssim \alpha \lesssim 1$, whereas a cosmic defect network gives a spectrum in the LISA band (for sufficiently large tension) scaling as a $plateau$ $\propto f^0$~\cite{Vachaspati:1984gt,Damour:2000wa,Damour:2001bk,Damour:2004kw,Fenu:2009qf,Figueroa:2012kw,Blanco-Pillado:2017oxo}. In certain inflationary scenarios, like in axion-inflation and its variants~\cite{Anber:2006xt,Barnaby:2010vf,Sorbo:2011rz,Pajer:2013fsa,Namba:2015gja,Ferreira:2015omg,Peloso:2016gqs,Domcke:2016bkh}, the spectrum of the GW signal may consist of a nearly flat part at low frequencies, followed by a smoothly growing part at high frequencies. 
The growing part of the signal could be reconstructed  as a power law for each different frequency bin, and this will be more and more accurate as long as the bin will be sufficiently small, and the signal has a large signal-to-noise ratio. On the other hand, cosmological sources like non-perturbative effects during post-inflationary preheating~\cite{Easther:2006gt,GarciaBellido:2007dg,GarciaBellido:2007af,Dufaux:2007pt,Dufaux:2008dn,Dufaux:2010cf,Figueroa:2017vfa,Adshead:2018doq} or strong first order phase transitions during the thermal era of the Universe~\cite{Kamionkowski:1993fg,Caprini:2007xq,Huber:2008hg,Hindmarsh:2013xza,Hindmarsh:2015qta,Caprini:2015zlo,Cutting:2018tjt}, typically generate a single or multi-peaked spectral signals. In such cases the final spectra within the LISA band might consist of one or more bumps. 

Our aim is to pursue a model-independent reconstruction
 of SGWB signals with distinct frequency dependence. For this purpose, we 
  have created  a catalog of benchmark signals exhibiting a variety of features, from single-slope power laws, to broken power laws, single- and double-peaked signals, and wiggly bumpy signals. The specific frequency dependence of our mock signals is given in Table~\ref{table:mock}. This is a representative choice of 
  frequency profiles, that  
   does not pretend to reproduce faithfully the predicted spectral shape of cosmological signals, but rather to mimic their basic features  which can also arise from 
   combining signals from several different sources. 
\begin{table}
\setlength{\extrarowheight}{2 ex}
\begin{tabular}{|l|l|l|l|}\hline
Class of the mock signal & Functional form of  $h^2\Omega_{\rm GW} \left( f \right) $ \\ [2 ex]\hline\hline
I. Single power law & $A_{0.001}\left( \frac{f}{0.001 \, {\rm Hz}} \right)^\gamma$ \\ [2ex] \hline 
II. Broken power law & $  A_{0.002}\left[\left({f\over0.002~{\rm Hz}}\right)^{\gamma}\Theta(f_T-f) + \left({f_T\over 0.002~{\rm Hz}}\right)^{\gamma}\left({f\over f_T}\right)^{\delta}\Theta(f-f_T)\right]$ \\ [2ex] \hline 
III. Single Peaked Signal: & $  A_b \, {\rm Exp}\left\lbrace -{[\log_{10}(f/f_b)]^2\over\Delta^2}\right\rbrace$ \\ [2ex] \hline 
IV. Double Peaked Signal: & $  A_1 \,{\rm Exp}\left\lbrace -{[\log_{10}(f/f_1)]^2\over\Delta_1^2}\right\rbrace + A_2 \,{\rm Exp}\left\lbrace -{[\log_{10}(f/f_2)]^2\over\Delta_2^2}\right\rbrace$ \\ [2ex] \hline 
V.  Wiggly Signal & $  A_w \, 10^{\sin\left(\Delta\log_{10}(f/f_w)\right)}$ \\ [2ex] \hline 
\end{tabular}
\caption{\it Different classes of mock signals studied in this work. }
\label{table:mock}
\end{table}

\subsection{Signal reconstruction}
\label{sec:reconstruction}

Here we apply the binning algorithm described in Section~\ref{sec:methodology} to analyze examples of signals belonging to the representative classes discussed in the previous section, and summarized in Table~\ref{table:mock}.
We make use of the {\tt SGWBinner} code to show the capabilities of the binning method to reconstruct these  benchmark signals, presenting some details of the reconstruction procedure. In each case, the total LISA frequency range is divided into frequency intervals. Within each interval (bin) we reconstruct the signal by means of a one-parameter or a two-parameter function, characterized by the signal amplitude in the first case, and the signal amplitude and slope in the second case. The code extracts the SNR of the reconstructed signal (that, for the examples considered here, turns out to be very close to the SNR of the injected signal). In all the examples studied in this section we assume an observation time of $4$ years with $75\%$ efficiency, meaning that we set $T = 3$ years.

In the majority of cases, we use the two-parameter binning procedure  described in section~\ref{sec_reconstruction}, reconstructing  both the amplitude and the slope of the signal within each bin. However, to demonstrate the many possible uses of our algorithm, in some cases we also show the results of a single-parameter binning procedure, reconstructing only the amplitude of the signal within each bin. We also present one example to show how this procedure 
allows to set an upper bound on the amplitude of a signal too small to be detected. 

As explained in section \ref{sec:methodology}, the code adaptively employs the AIC to establish whether it is favourable to merge nearby bins. If, at the end of this procedure, only one bin remains in the full LISA frequency band, this means that it would have been equally convenient to fit one single power law to the data in the first place. To demonstrate that the binning procedure we developed does indeed improve the fit to complicated signals with respect to the single power law assumption, for each example under analysis in this section we calculate the difference between the AIC of the reconstruction \texttt{SGWBinner} performs, and the AIC of a single power law reconstruction in the full LISA frequency band. This latter is evaluated by fixing the initial number of bins to one and turning off the adaptive bin-size procedure. In all examples, besides obviously the single power law one of subsection \ref{sec:PLS_reconstruction}, the AIC of the full reconstruction is manifestly smaller than the single power law reconstruction. This demonstrates the advantage of using \texttt{SGWBinner} with respect to previous data analysis techniques based on single/broken power laws: it performs equally well in these last cases, but it performs better for more complicated spectral shapes.

\subsubsection{Single power law signal}
\label{sec:PLS_reconstruction}

\begin{figure}[t!]	\includegraphics[width=0.52\textwidth]{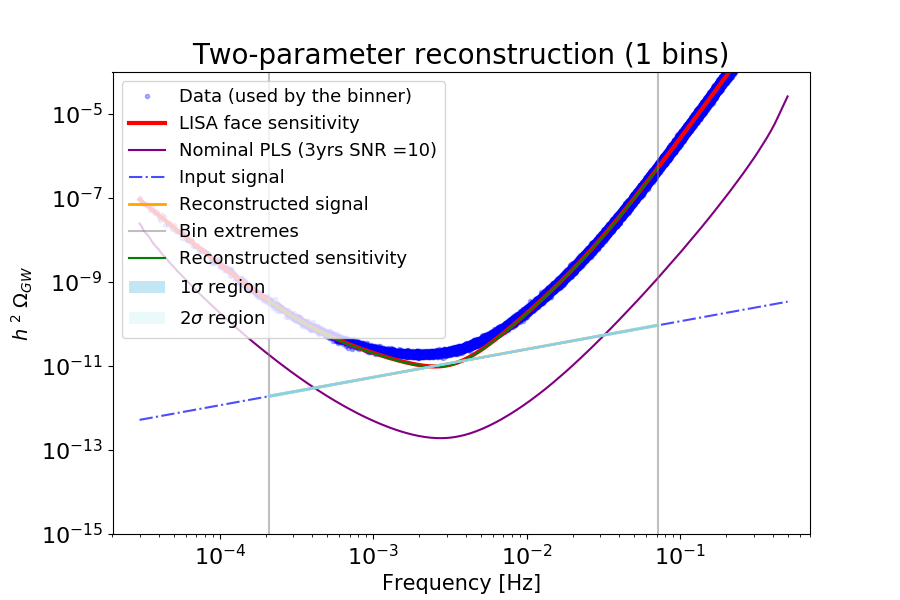}
	\includegraphics[width=0.52\textwidth]{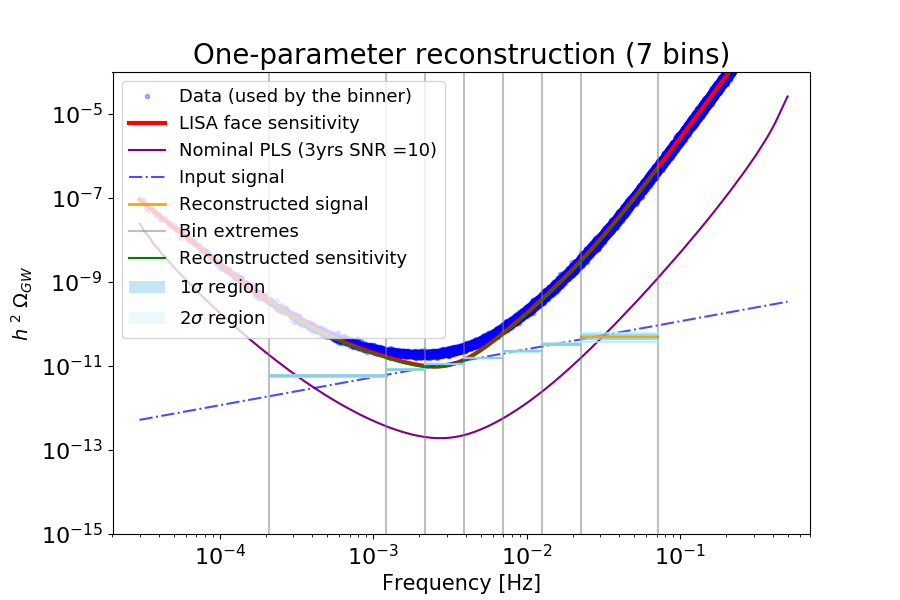}
	\caption{\it Left panel: two-parameter reconstruction with $N_{\rm bins} = 5$ initial bins. Right panel: one-parameter reconstruction with $N_{\rm bins} = 10$ initial bins.  Note that in the left panel the error bar line is thin and superimposed on the input signal, therefore the latter is not visible.}
\label{fig:1}
\end{figure}

The simplest benchmark signal to reconstruct is a single power law, given in the first line of Table \ref{table:mock}. We choose an input signal with amplitude $A_{0.001} = 5.4 \cdot 10^{-12}$ and slope  $\gamma = 2/3$, motivated by the astrophysical SGWB from BH and NS binaries \cite{LIGOScientific:2019vic}. The results of the reconstruction algorithm are presented in the left and right panels of fig.~\ref{fig:1}, for which we have used, respectively, two signal parameters (amplitude and slope) and one signal parameter (amplitude only) per bin. The SNR of the reconstructed signal with respect to the reconstructed noise curve is 601.

For the two-parameter reconstruction, the reconstructed signal (orange line) matches the input signal very well within the error bars (associated with the thickness of the light-blue curves -- see also Section \ref{sec:implementation} for a description of the conventions we use to visualize the results). The input signal is a single power law across the whole LISA band: therefore, for a fit based on two parameters per bin, a successful reconstruction should converge to a single bin. The adaptive bin-merging procedure based on the AIC indeed results in one final bin (c.f. the left panel of fig.~\ref{fig:1}). We have started with $N_{\rm bin}=5$ initial bins, but verified that the merging to one final bin occurs also for larger $N_{\rm bin}$. 

For the one-parameter fit, the required number of both initial and final bins is larger (c.f. the right panel of fig.~\ref{fig:1}): we have verified that an accurate reconstruction needs at least ten initial bins. Note that in the outermost left and right bins the input signal is far off the error bars of the reconstructed one: this is because the fit is dominated by the part of the input signal with the highest SNR within the bin, while the SNR of the input signal is too low to justify further bin splitting. This feature is visible in many of the examples that follow, meaning that the reconstructed signal and the associated error bars cannot be trusted when the former extends well below the PLS.

\begin{figure}[h!]
\centering	\includegraphics[width=0.7\textwidth]{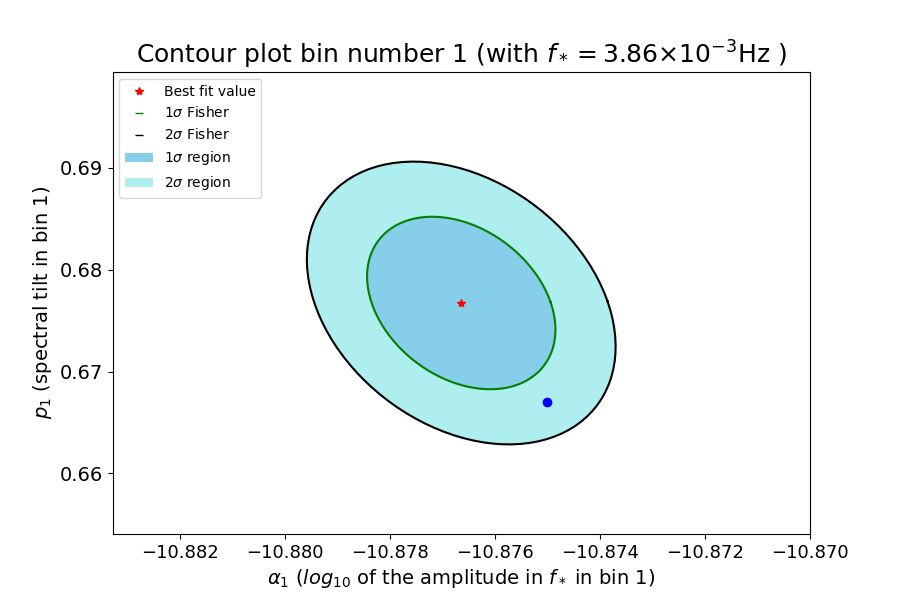}\caption{\it Best fit value (red dot), and 1$\sigma$ and 2$\sigma$ contour lines (green and black lines) for the amplitude and slope of the reconstructed signal in the final bin (c.f.~the left panel of fig.~\ref{fig:1}).  The blue dot represents the signal input values. }
	\label{fig:contourPL}
\end{figure}

Fig.~\ref{fig:contourPL} shows the 1$\sigma$ and 2$\sigma$ contours of the amplitude and slope of the reconstructed signal in the final bin, together with their best fit values (c.f.~section~\ref{sec:implementation}).  The log$_{10}$ of these latter is $\left\{-10.876,\, 0.676 \right\}$, while the log$_{10}$ of the signal input values is $\left\{-10.875,\, 0.667 \right\}$ (blue point in fig.~\ref{fig:contourPL})~\footnote{The input signal was generated at pivot frequency $f_* = 1$ {\rm mHz} with $A = 5.4 \cdot 10^{-12}$. Here we recalculate the input amplitude at the new pivot frequency used by the code to compute the best fit.}. The 1$\sigma$ and 2$\sigma$ contours computed using the marginalized likelihood (blue-shaded regions) and the Fisher forecast (green and black lines) are in very good agreement.

\subsubsection{Broken power law signal}
\label{sec:general_reconstruction}

\begin{figure}[h!]%\centering
\includegraphics[width=0.52\textwidth]{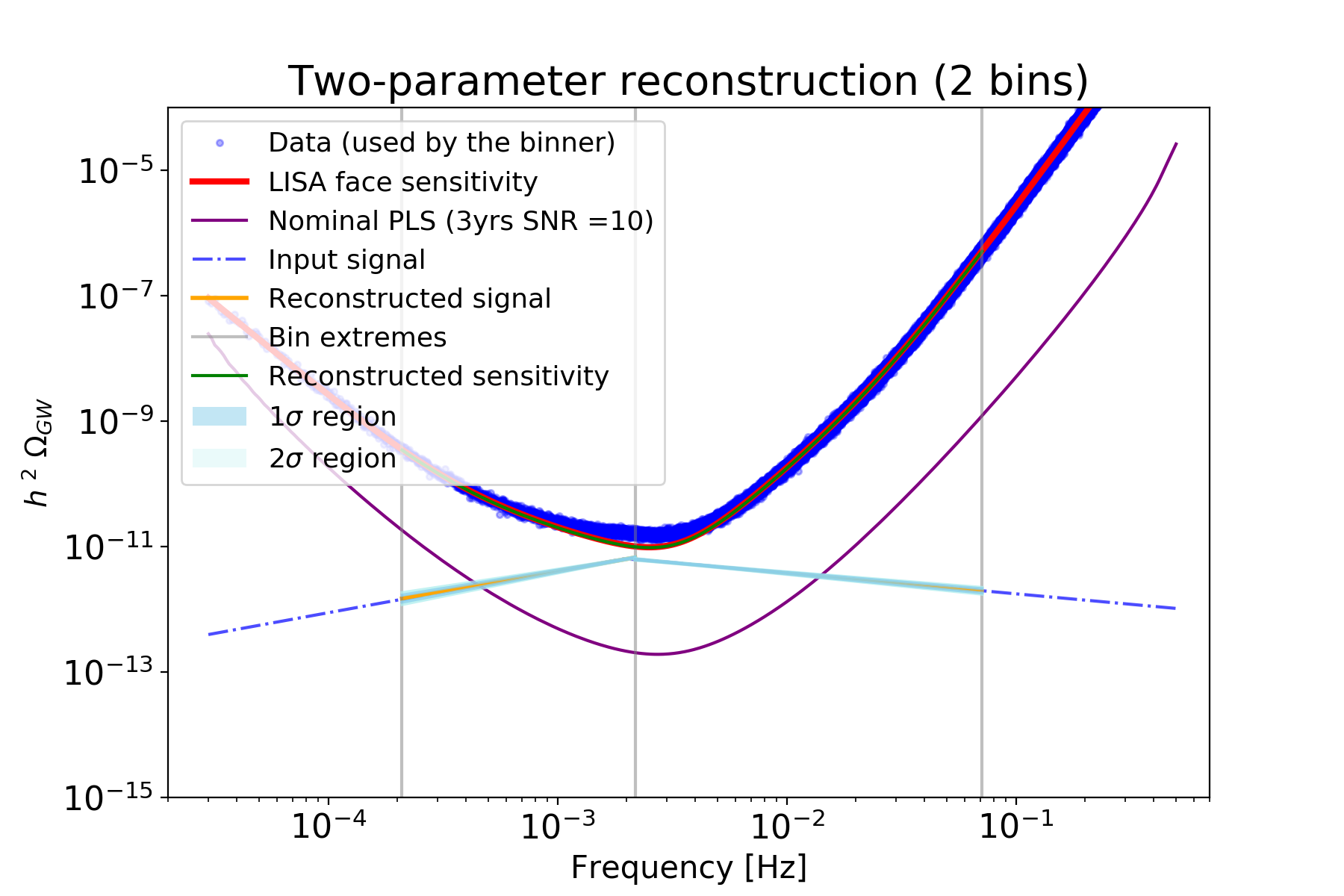}~	\includegraphics[width=0.52\textwidth]{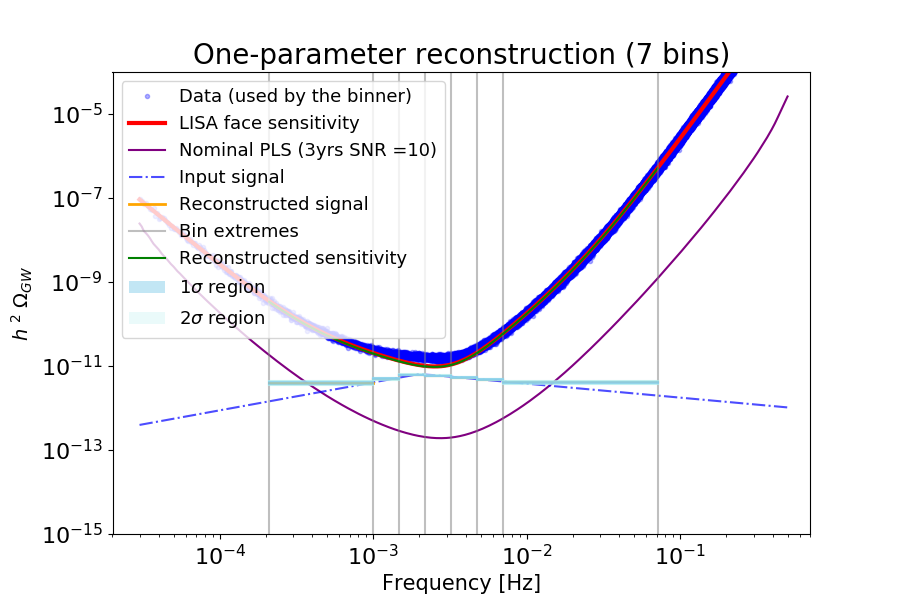}\caption{\it Left panel: two-parameter reconstruction with $N_{\rm bins} = 5$ initial bins. Right panel: one-parameter reconstruction with $N_{\rm bins} = 15$ initial bins.}
	\label{fig:broken}
\end{figure}

Here we consider a {broken power law} signal: a piece-wise linear function in the plane $\{ \log f ,\log(h^2 \Omega_{\rm GW }) \}$, changing slope at some given frequency. Specifically, we choose the functional form given in the second line of Table \ref{table:mock}, with parameters $A_{0.002} = 6.5 \cdot 10^{-12}$, $\gamma= 2/3 $, $\delta= - 1/3 $ and $f_{{\rm{T}}}=0.002$. This spectral shape might arise from the combination of two physically distinct sources, and it is represented with a blue dot-dashed line in fig.~\ref{fig:broken}.

The left panel of fig.~\ref{fig:broken} shows the signal reconstruction with two parameters per bin. The SNR of the reconstructed signal is about $ 330$. We expect an efficient reconstruction to converge to two final bins, separated at a frequency close to the input value $f_T=0.002$. Indeed, the result is two bins with $0.00021 \leq  f/{\rm Hz} \leq 0.0021$ and $0.0021 \leq f/{\rm Hz} \leq 0.071$ respectively.

\begin{figure}[t!]	\includegraphics[width=0.52\textwidth]{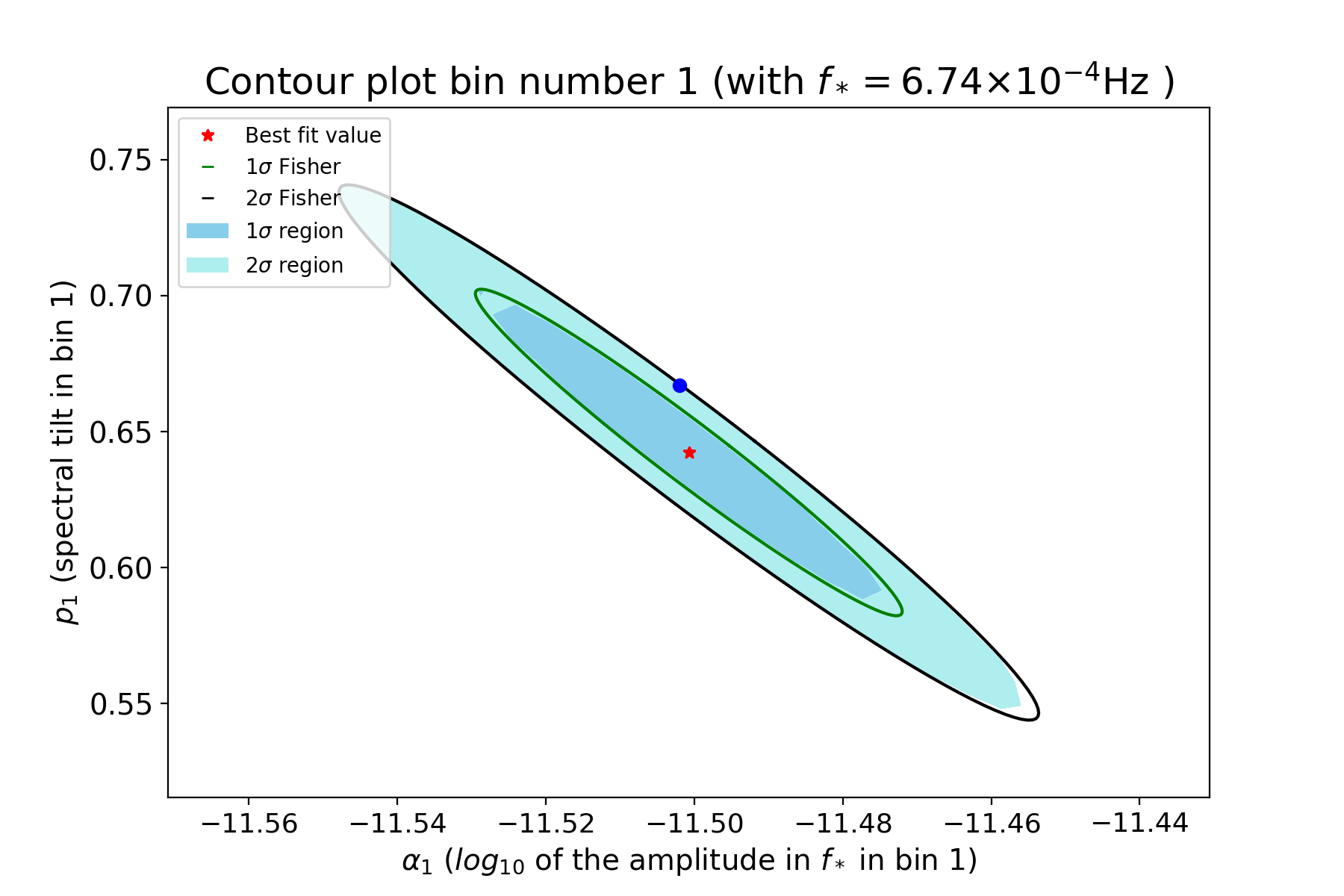} 
\includegraphics[width=0.52\textwidth]{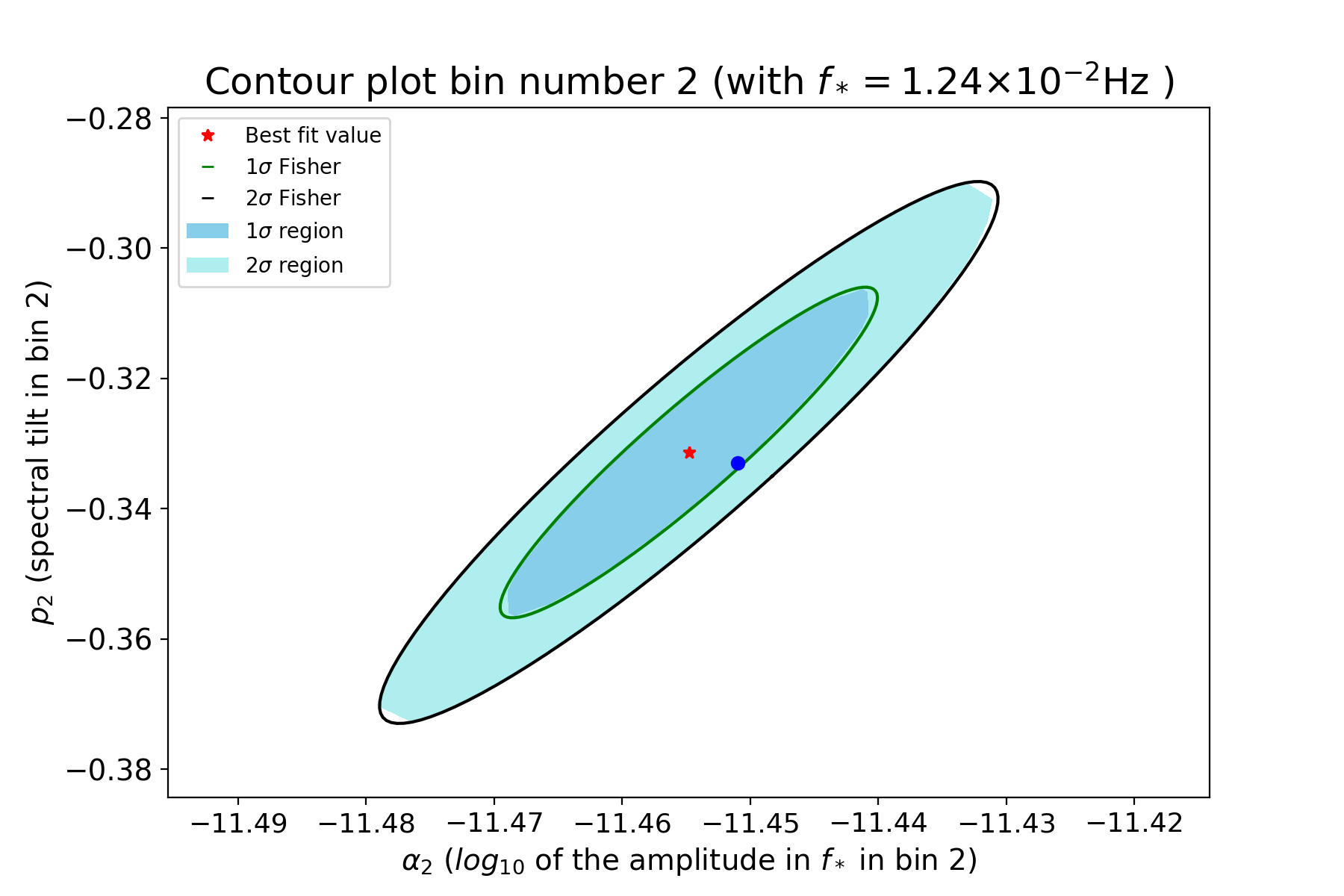}
	\caption{\it Best fit value, and 1$\sigma$ and 2$\sigma$ contours for the amplitude and slope of the reconstructed signal in the two bins of the left panel of fig.~\ref{fig:broken}. The blue dots represent the signal input values: for more details, c.f.~the main text. 
	}\label{fig:4}
\end{figure}

Since the two bins overlap with the regions where the input signal is a single power law, we can easily compare the input and reconstructed signals by writing the former in the form (\ref{2-param}) within each bin. The input parameters are $\left\{ \alpha ,\, p \right\} = \left\{ -11.502, 0.66 \right\}$ in the first region, to be compared with the best-fit values $\left\{-11.500 ,\, 0.64 \right\}$ for the first bin; and analogously, $\left\{ \alpha ,\, p \right\} = \left\{ -11.451, -0.33 \right\}$ for the second region, to be compared with the best-fit values $\left\{-11.454 ,\, -0.33 \right\}$ for the second bin. These values, together with the 1$\sigma$ and 2$\sigma$ likelihood and Fisher matrix contours, are shown in fig.~\ref{fig:4}.

The input signal being more complex than a single power law, in this case we do expect the algorithm of the \texttt{SGWBinner} code to provide a better fit than a single power law fit in the whole LISA frequency band (contrary to the example of subsection \ref{sec:PLS_reconstruction}, for which no improvement is expected). We have therefore calculated the AIC for the two fits (c.f.~the preamble of section \ref{sec:reconstruction}): indeed, the multi power law reconstruction leads to an improvement in the AIC of $\Delta {\rm AIC}\equiv {\rm AIC_{multi PL}}- {\rm AIC_{PL}} \simeq -\,6\,\cdot 10^2$ with respect to a single power law reconstruction.

The one-parameter signal reconstruction is shown in the right panel of fig.~\ref{fig:broken}. We performed several runs with different numbers of initial bins, and we  found that the optimal one (balancing running time and final result) is $N_{\rm bins} = 15$. The reconstruction naturally requires greater numbers both of initial and of final bins with respect to the two-parameter one. The difference between the AIC for the binned one-parameter reconstruction and the AIC for the single power law reconstruction is about $-6\cdot 10^2$.

\subsubsection{Single peaked signal}

Certain sources (see e.g.~\cite{Namba:2015gja}) produce SGWB signals amplified at well defined frequencies in the LISA band, well described by the functional form in the third line of Table  \ref{table:mock}. Here we consider an input signal with a bump in the central part of the LISA frequency band, with amplitude $A_b = 10^{-11}$, central frequency $f_b = 0.003 \, {\rm Hz}$, and width $\Delta = 0.2$ (parametrised as in Table  \ref{table:mock}). 
 
\begin{figure}[t!]\centering
\includegraphics[width=0.60\textwidth]{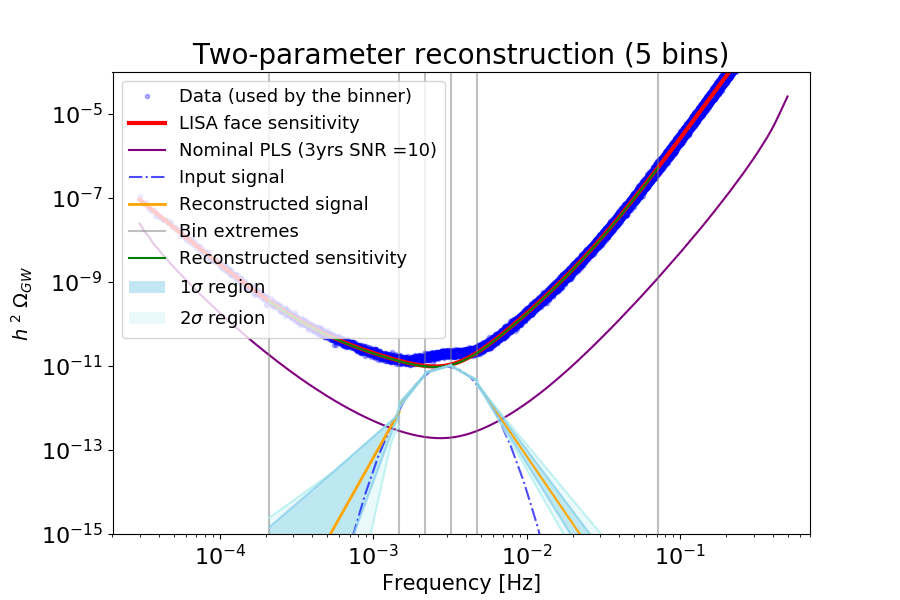}
	\caption{\it Two-parameter reconstruction of a signal with a bump, fixing $N_{bins} = 20$.}\label{fig:5}
\end{figure}

As shown in fig. \ref{fig:5}, the reconstruction is accurate in the central part of the frequency band, where the signal peaks, and worsens at the extrema. Note that the number of initial bins must be sufficiently large for a good reconstruction, $N_{\rm bins} \geq 20$. With respect to the reconstructed noise curve,
the signal has a total SNR $ \simeq  360$. 
Comparing to a single power law reconstruction, the improvement in the AIC is $\Delta {\rm AIC} \simeq -2\cdot 10^4$.

\subsubsection{Double peaked signal}

Two distinct SGWB sources could provide a signal with two peaks at different frequencies in the LISA band. The functional form of the signal is given in the fourth line of Table~\ref{table:mock}, and here we choose the parameter values as follows:  $A_{1} = 2.0 \cdot 10^{-10}$, central frequency $f_1 = 7 \cdot 10^{-4}$ Hz, $\Delta_1 = 0.25$; $A_{2} = 2.0 \cdot 10^{-10}$, central frequency $f_2 = 0.02$ Hz,   and $\Delta_2 = 0.25$. 

\begin{figure}[h!]\centering
\includegraphics[width=0.60\textwidth]{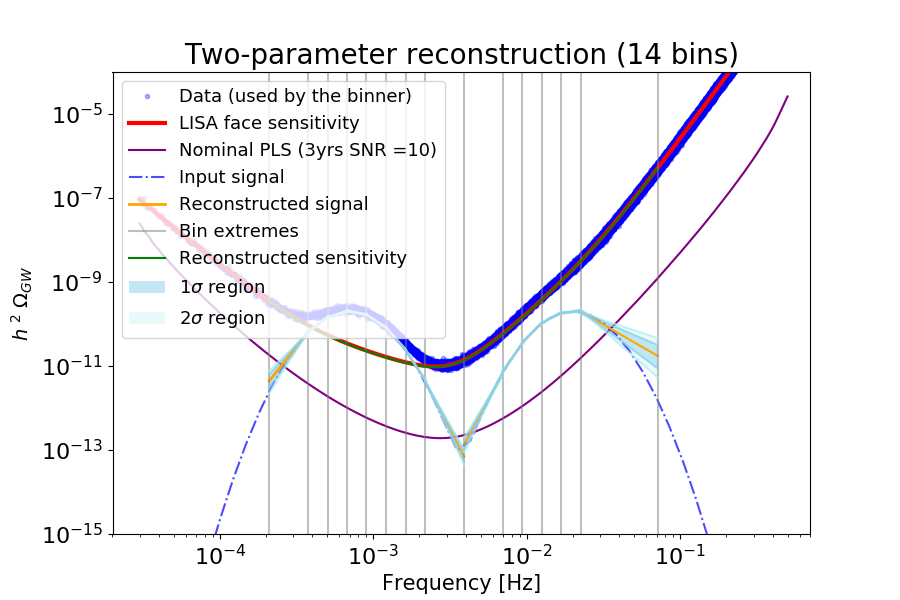}
\caption{\it Two-parameter reconstruction of a signal with two peaks, with $N_{\rm bins} = 20$.}\label{fig:6}
\end{figure}

Figure~\ref{fig:6} shows a very good signal reconstruction of both signal peaks, performed via the two-parameter fit. The AIC method is again effective to reduce the number of useful bins. However, as for the single peak case, a large number of initial bins is required to better capture the features of this kind of signals, given their fast variation with frequency: here we have again used $N_{\rm bins} = 20$. The improvement in the fit with respect to the single power law reconstruction is also similar: $\Delta {\rm AIC} \simeq -10^5$. The reconstructed signal results to have SNR $\simeq  1.7\cdot 10^3$.

The two bumps could be separated by a hole: we have tested the ability of the {\tt SGWBinner} code to reconstruct the signal also in this scenario. 
The functional form of the signal is again the one given in the fourth line of Table~\ref{table:mock}, with the following parameters:  $A_{1} = 10^{-16}$, central frequency $f_1 = 0.001$ Hz,  and $\Delta_1 = 0.3$; $A_{2} = 10^{-16}$, central frequency $f_2 = 0.01$ Hz,   and $\Delta_2 = 0.3$. 

\begin{figure}[h!]\centering
\includegraphics[width=0.49\textwidth]{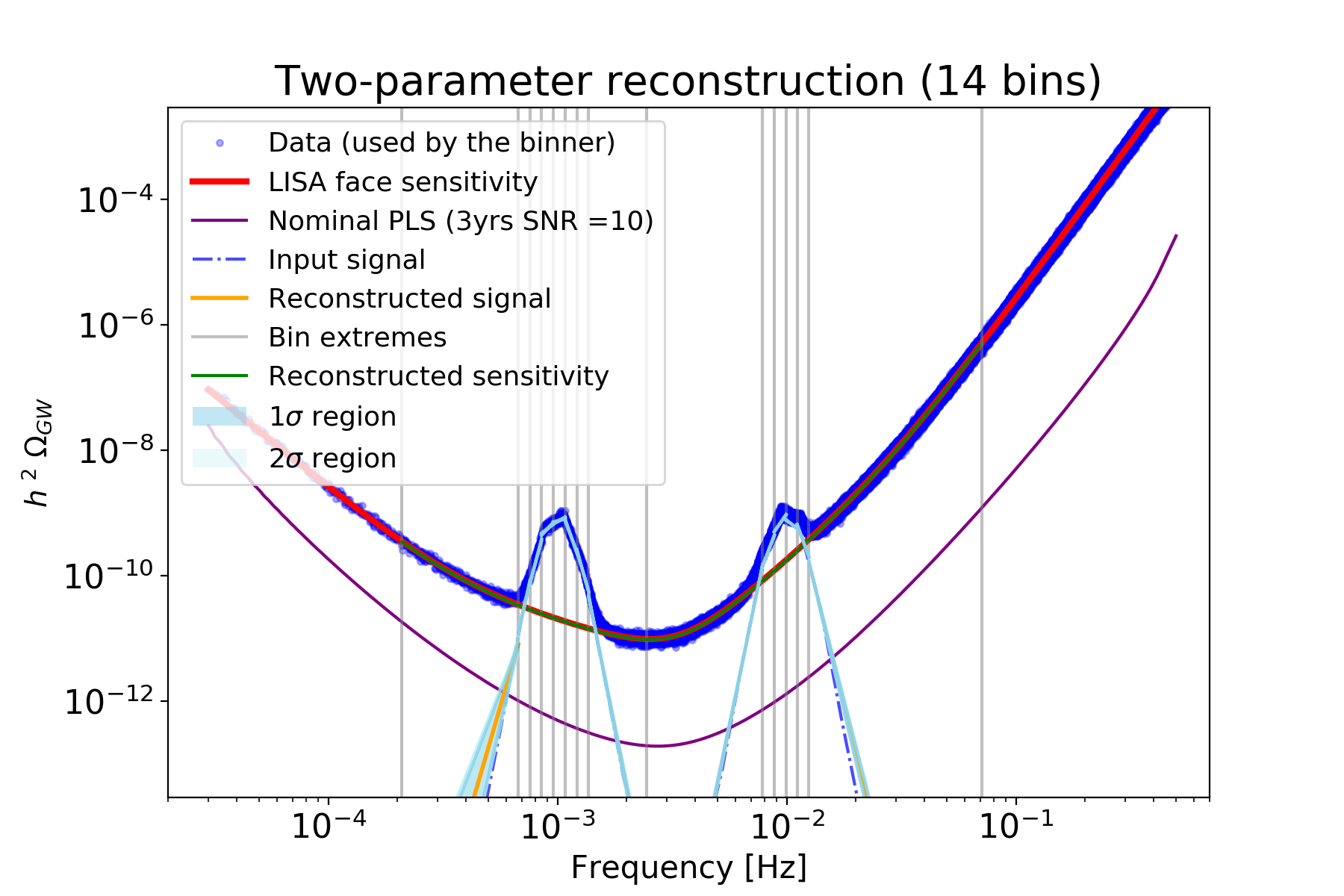}
\includegraphics[width=0.49\textwidth]{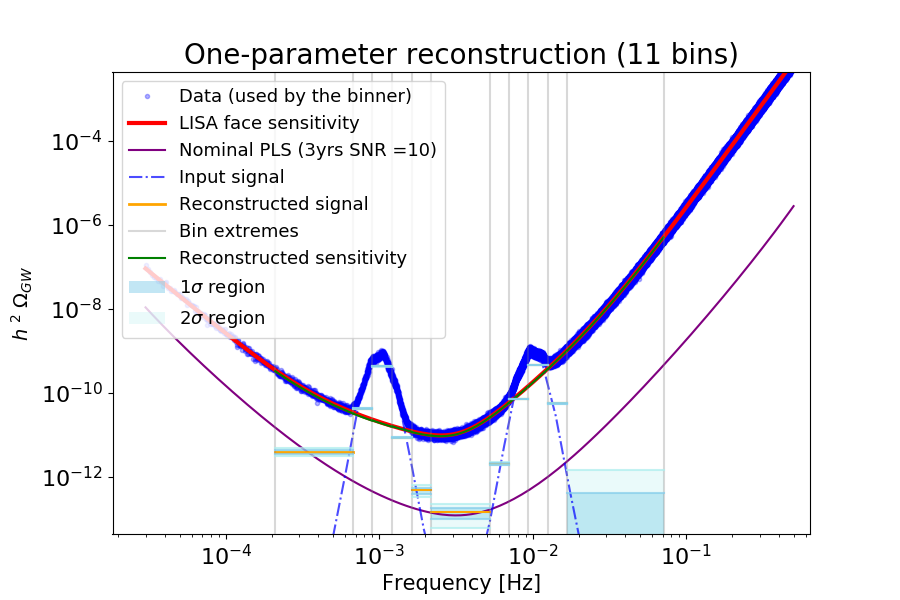}
	\caption{\it Two-parameter reconstruction (left panel) and one-parameter reconstruction (right panel) of a signal with a hole in the middle of the LISA sensitivity band. The signal has been analyzed with $N_{\rm bins} = 50$ for the two parameter case and $N_{\rm bins} = 30$ for the one parameter case. 
	}\label{fig:7}
\end{figure}

The left panel of fig.~\ref{fig:7} shows the two-parameter -- amplitude and slope -- fit of this signal, while the right one the one parameter -- amplitude-only -- fit. In both cases the reconstruction is significantly more accurate than the single power law one: the improvement in the {\rm AIC} is $\simeq -\, 4 \cdot 10^5$. However, a large number of initial bins is necessary for the reconstruction: we have used 50 and 30 initial bins for the two-parameter and one-parameter reconstructions respectively. The  SNR of the reconstructed signal is about $6.4 \cdot 10^3$. 

For this benchmark signal we have also performed the one-parameter fit, to illustrate the reconstruction procedure when the signal is well below the PLS (right panel of fig.~\ref{fig:7}). In this analysis the signal in the bin at $2\cdot 10^{-2}\,{\rm Hz}\lesssim f\lesssim 7\cdot 10^{-2}\,{\rm Hz}$ is compatible with the noise-only hypothesis and, as expected, its 2$\sigma$ upper bound is below the PLS.

\subsubsection{Wiggly signal}

\begin{figure}[t!]
\centering \includegraphics[width=0.48\textwidth]{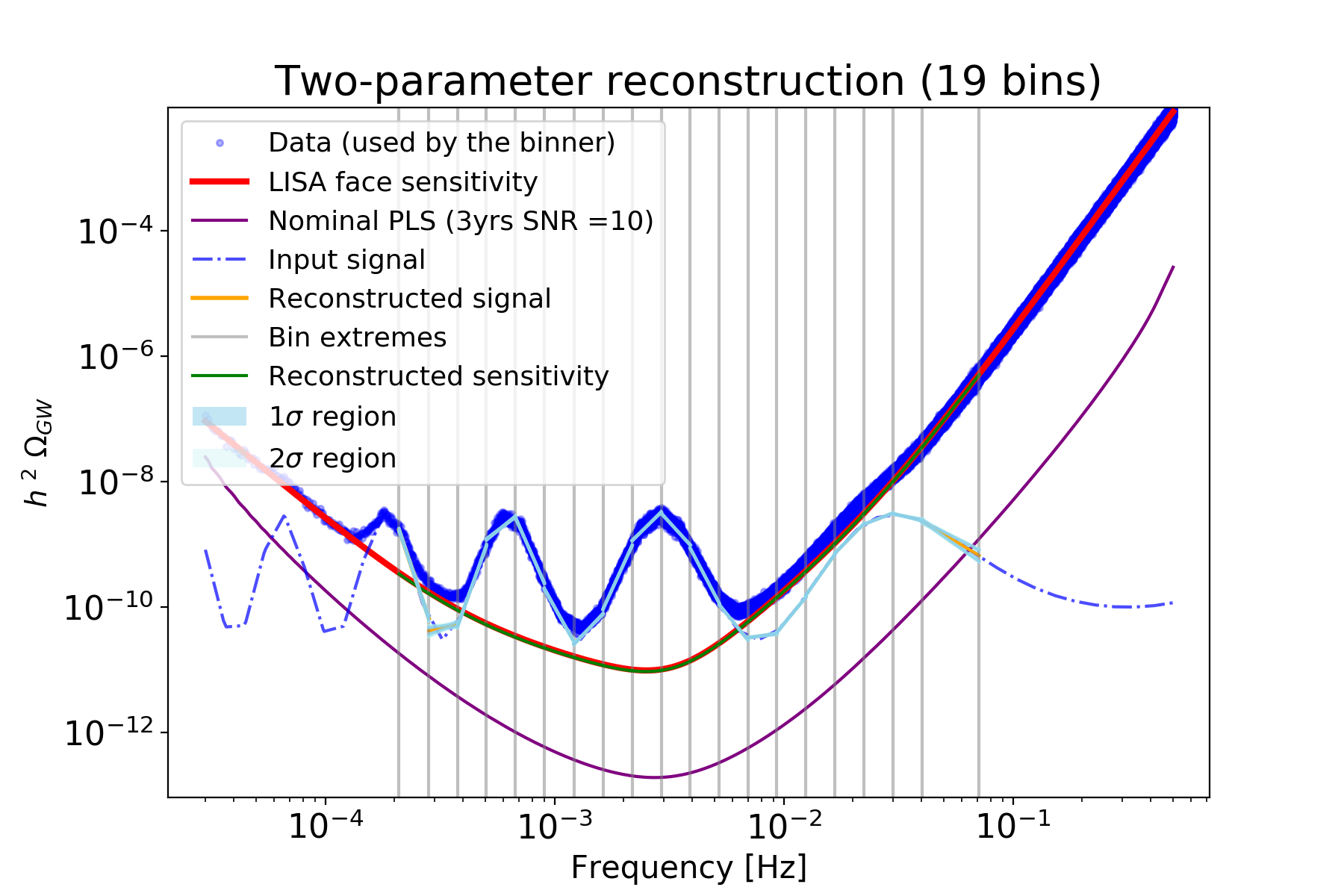}	
\includegraphics[width=0.48\textwidth]{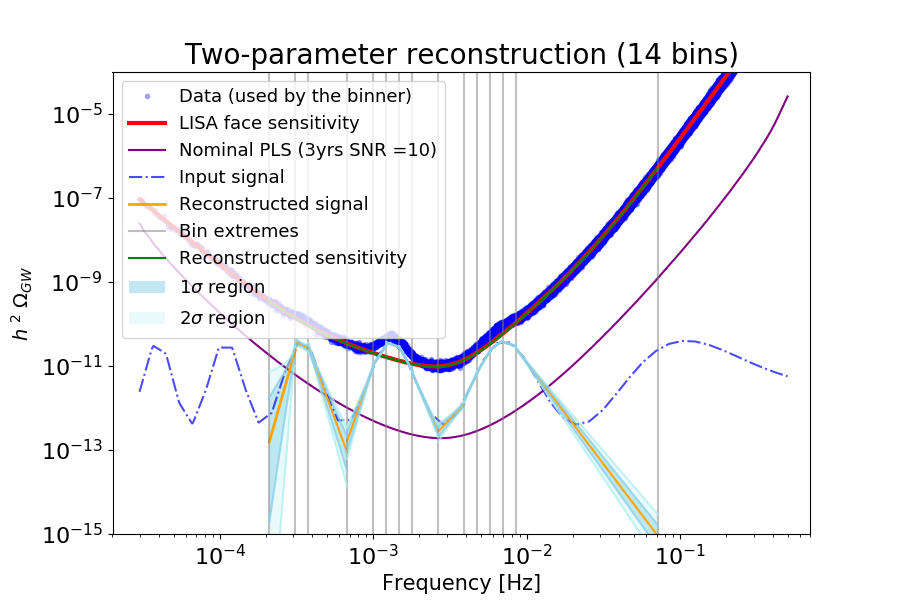}
 \caption{\it Two-parameter reconstruction of two different wiggly signals. The signal in the left panel has been analyzed with $N_{\rm bins} = 20$ initial bins, while the one on the right, with lower signal amplitude, has been analyzed with $N_{\rm bins} = 30$.}\label{fig:8}
\end{figure}

Here we discuss a wiggly signal, which we analyse as a toy model to test the reconstruction procedure in an extreme case. The functional form of the signal is given in the fifth line of Table \ref{table:mock}. We consider two signals with two different sets of values for the input parameters: $A_{w} = 3.0 \cdot 10^{-10}$, $\Delta = 2.0$ and $f_w = 0.1$ {\rm {Hz}} (left panel of fig.~\ref{fig:8}); $A_{w} = 4.0 \cdot 10^{-12}$, $\Delta = 1.7$ and $f_w = 0.1$ {\rm {Hz}} (right panel of fig.~\ref{fig:8}). 

The signal in the left panel of fig.~\ref{fig:8} has a larger amplitude (SNR $\simeq 80000$) than the one in the right panel (SNR$\simeq  575$). Not surprisingly, a visual comparison of the two panels shows that the first signal is reconstructed better than the second one, particularly in the outermost regions of the reconstruction frequency domain. For both signals, the multi power law largely improves the fit compared to the single power law; in particular for the higher amplitude signal $\Delta {\rm AIC} \simeq -7 \cdot 10^{5} $, while for the lower amplitude signal $\Delta {\rm AIC} \simeq -\,9 \cdot 10^{4}$.

We have analysed this particular signal to emphasize the ability of the code to reconstruct SGWB shapes beyond the standard, predicted ones. This opens up the possibility to detect even potentially new SGWB sources. The analysis shows that the signal can be reconstructed very well in regions of the LISA frequency band where it has a sufficiently large SNR.

\smallskip

\section{SNR threshold}
\label{sec:SNR_thr}

In this section we investigate the lowest SNR value for  a  power law signal, in order to be detectable at LISA.
In particular, we aim at defining the threshold SNR$_{\rm thr}$ 
used in section~\ref{sec:SGWB_at_LISA} for the construction of the PLS. 

\begin{figure}[t!]\centering
    \includegraphics[width=0.72\textwidth]{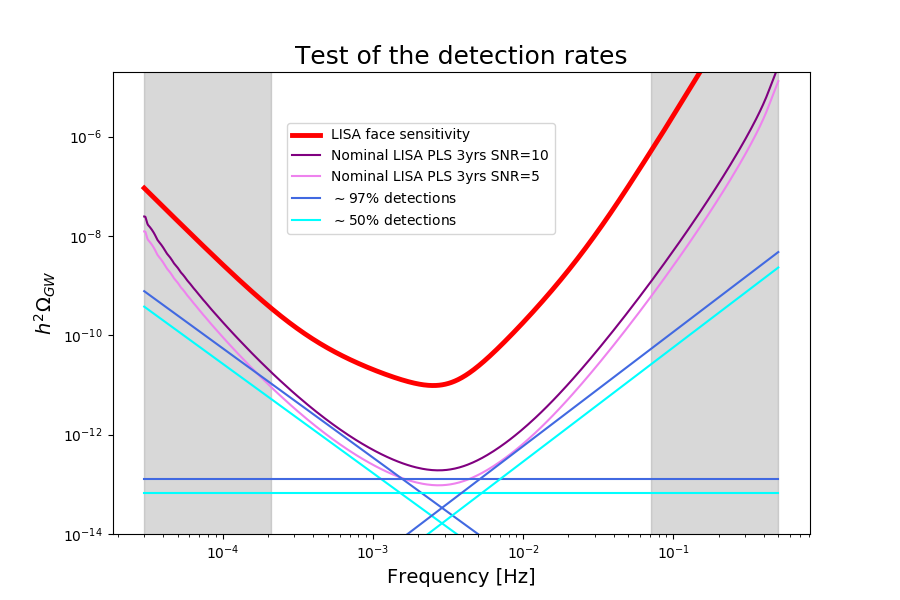}
    \caption{\it The sensitivity curve derived from the noise model we adopt (red curve) and
    the corresponding PLS for $T=3\,$yr with SNR$\,=5$ (purple curve) 
and SNR$\,=10$ (violet curve), compared to some power law signals which, 
out of 100 realizations, are detected about 97\% (dark blue lines) and 50\% 
(light blue lines) of the times. The gray areas are the regions we previously used to infer the noise prior.}\label{fig:snr_thr}
\end{figure}

Note that the construction of the PLS assumes the same SNR$_{\rm thr}$ 
for all power law slopes.
One may wonder whether this property holds in practice. In fact, different signal  
slopes correspond to different frequency regions where the integral defining the SNR is dominated. 
For some particular slopes, one might have to integrate on a frequency region where the signal is very similar to the noise. 
Consequently, two power law signals with the same SNR might not be equally compatible with the noise-only hypothesis (with some offsets in the 
noise parameters). In this case, for the construction of the PLS one would have to choose a higher SNR$_{\rm thr}$ for 
those power laws more easily  confused with a noise contribution. 
This is the reason why the value of SNR$_{\rm thr}$ may depend on the slope. 
On the other hand, here we show that it is reasonable to
use the same value, SNR$_{\rm thr} \simeq 10$, 
at least for all the power law curves tangent to the PLS in the frequency band  between the grey regions of  
fig.~\ref{fig:snr_thr}.

For this task we set \texttt{SGWBinner} on a global fit  option, so that the code skips the binning procedure, and fits the 
data in the whole LISA frequency band. The data in the outermost regions of the frequency band (grey  
areas in fig.~\ref{fig:snr_thr}) are thus {\it not} used to infer the noise 
prior (as done in the analyses of the previous sections), but are instead 
included in the fit. We focus on three families of 
parallel power laws:
\begin{equation}
h^2\Omega_{\rm GW}(f)= 10^{\alpha} \left(f/\sqrt{3 {\rm \,mHz}}\right)^{p}~,
\label{input-SNR}
\end{equation}
with $p=-2.2,0,2.3$ and $\alpha \in \{-15,-14.92, -14.84, \dots , -11\}$. 
For 
each pair of $\alpha$ and $p$, we generate 100 realizations of signal 
plus noise data (with $T=3\,$yr, $A=3$ and $P=15$), we run 
\texttt{SGWBinner}, and we count how many reconstructions out of 100 are 
incompatible (at 2$\sigma$ level)  with the noise-only  
hypothesis~\footnote{Note that, in practice, the code does not  explore the $\alpha = 
\log \left( {\rm amplitude} \right)  = - \infty$ value. For this reason, 
for this study, we declare that an analysis is compatible with a 
noise-only hypothesis if the value $\alpha = - 40$ is in the $2 
\sigma$ output contour interval. We verified that this arbitrary 
threshold can equivalently be replaced with any other  value well below 
the PLS curve, since (as expected) the likelihood flattens  in this 
region.}.

Figure~\ref{fig:snr_thr} displays the result of this analysis. The dark blue 
(respectively, light blue) straight lines are the power law signals that 
turn out to be incompatible with the noise-only hypothesis in around 97 
(respectively, 50) out of 100 realizations. This implies that a power 
law signal parallel and above one of the dark blue (respectively, light blue) 
lines will be detected with a probability greater than $97\%$ 
(respectively, $50\%$). The dark blue, horizontal line, corresponding to an 
input signal of the form (\ref{input-SNR}), with $\alpha_1\simeq-12.9$ 
and $p=0$,
is slightly below the PLS calculated for SNR$_{\rm thr}=10$ and 
$T=3\,$yr (purple curve) and marginally crosses the PLS for SNR$_{\rm 
thr}=5$ and $T=3\,$yr (violet curve). The SNR of this line is in fact 
around 7, which confirms that the choice SNR$_{\rm thr} \simeq 10$ for a 
flat SGWB signal is a reasonable one, as found also in \cite{Adams:2013qma}. 
This indicates that the \texttt{SGWBinner} procedure is reliable, being compatible with 
the previous result of~\cite{Adams:2013qma}. 
Furthermore, we show that SNR$_{\rm thr}$ can be interpreted as the probability of 
detecting a signal. This statistical interpretation implies that, with some luck, a signal 
with SNR$ < $SNR$_{\rm thr}$ can also be detected. For instance, a 
flat signal with $\alpha\simeq -13.2$ (light blue, horizontal line in 
fig.~\ref{fig:snr_thr}) can be detected $\sim$50\% of the times.

We obtain reasonable results also for the other two families of power laws under analysis. 
The dark blue lines with $p=-2.2$ and $p=2.3$ in fig.~\ref{fig:snr_thr} (detectable 
$\sim$97\% of the times) are tangent to the PLSs with SNR$_{\rm thr} \simeq 10$ and SNR$_{\rm 
thr} \simeq 5$, respectively.
This proves that the SNR$_{\rm thr}$ does indeed depend on the slope, but it varies only within a factor of 2 in the interval $5\times 
10^{-4}$\,Hz$~\lesssim f \lesssim 1\times 10^{-2}\,$Hz. Therefore, the 
choice SNR$_{\rm thr}\simeq 10$ is reasonable to construct the PLS in the frequency region between the gray bands~in fig.~\ref{fig:snr_thr}.  

On the other hand, for frequencies $f\lesssim 10^{-4}$\,Hz and $f\gtrsim 
10^{-1}$\,Hz the LISA sensitivity curve is well approximated by two power laws. 
This hints to the fact that a larger SNR$_{\rm thr}$ might be required for power law 
signals parallel to the sensitivity curve in this regions. This 
is what some preliminary results (not shown here) suggest. 
However, since the PLS is intended to be a qualitative, graphical tool, minor 
dependencies on the value of SNR$_{\rm thr}$ do not practically impact its utility.

\section{Conclusions}
\label{sec_conclusions}

In this work we presented an algorithm at the basis of  the \texttt{Python3} code    \texttt{SGWBinner},  developed for analyzing and reconstructing   stochastic gravitational wave backgrounds (SGWBs)
that can be detected  by LISA.  This effort is motivated by the fact that many different gravitational wave (GW) sources can  contribute in forming  a SGWB characterized by a complex frequency  profile.  The data are divided in a number of bins, which is dynamically determined by our algorithm according to the Akaike Information Criterion. In each bin, the 
data are fitted by either a constant amplitude, or a single power law (depending on the choice of the user). The algorithm is therefore agnostic on the underlying signal, and it can be used as a ``first pass'' in the data analysis. Once it has been employed using the \texttt{SGWBinner}, and once some features   emerge from the data, one can then use ad-hoc, theoretically motivated  templates for reconstructing more faithfully the signal profile.   
We have also proven that the algorithm is preferable to  SGWB searches based uniquely on power law or broken power law templates~\footnote{Specifically, for  benchmark scenarios with complex frequency shapes, the algorithm always achieves a lower value of the Akaike Information Criterion than the one inferred using a power law template. When the input signal is a power-law or a broken power-law, the algorithm reconstructs the signal as well as the searches based on the corresponding templates.}.

Beside  theoretical models, the code also simulates  the  optical metrology system  noise PSD (dominant at high frequency) and the mass acceleration noise (dominant at low frequency), with a functional dependence given in eqs. (\ref{eq:noise-AB}). The noise model introduces two parameters $A$ and $P$, that the \texttt{SGWBinner} reconstructs bin-by-bin. The external regions of the LISA frequency band are used to set priors on these parameters (the width of these regions in frequency space can be selected by the user). After analysing the data, the user is left with  a set of best fit values $\left\{ A_i \right\}$ and  $\left\{ P_i \right\}$ for these two parameters, one from each bin. If these values are compatible with each other, then one can conclude that the functional dependences given in eqs. (\ref{eq:noise-AB}) are a good model of the LISA noise. If this is not the case, a different noise model should  be employed, and the {\tt{SGWBinner}} can be easily modified to work with it.   As the external regions are used to set priors on the noise parameters, we believe that extending the LISA band while still having a good modelling of the noise, will help to better reconstruct the noise curve and, in turn, the signal. 

In simulating the signal and the noise data, we assumed a set of chunks of data of around 11 days each (which is an estimation for the amount of time for which the instrument can provide continuous measurements  \cite{Audley:2017drz}). We made the hypothesis that the signal and the noise remain stationary across all the chunks, so that the  \texttt{SGWBinner} treats each chunk as an independent realization drawn from the same statistics. It is in principle straightforward to test whether the actual data respect this assumption, and (if needed)  to perform a more sophisticated analysis, where  the noise model is allowed to vary between chunks, or the signal varies with time (this will be for example the case if the stochastic GW background is not isotropic).  The best way to concretely     investigate these technical aspects  will be to implement our procedure  in the LISA pipelines, by linking \texttt{SGWBinner} to the LISA simulator.  This will be  necessary to  take into full account the evolving state of the art on LISA instrumental performances.

The
users can employ \texttt{SGWBinner} to forecast the reconstruction prospects of a given signal~\footnote{Information about the code can be obtained by contacting the corresponding author of this paper. We plan to make the code public (likely in early 2020) after further developments and applications within the LISA Consortium.}.
 However,  some insights are already possible by means of the power-law sensitivity (PLS) and binned PLS tools we have provided. The former allows one to assess whether a SGWB signal is in the ballpark of detection. The latter whether some substructure of the frequency shape can be  reconstructed. Both curves rely on the SNR$_{\rm thr}$ condition, which is a pragmatic criterion to infer whether a power law can be reconstructed in a given frequency interval. We demonstrated that a reasonable  choice for the SNR is SNR$_{\rm thr}\simeq 10$,  given the qualitative purposes of the PLS and binned PLS utilities.

To conclude, we have presented an algorithm that allows for  analysizing   SGWB profiles without limiting  assumptions on the signal frequency shape. We have shown that the LISA experiment has the potential to reconstruct complex SGWB frequency profiles; on the other hand it would be interesting to apply our  data analysis strategy to other detectors, covering different
frequency bands of GW signals. 
This promising approach is of paramount importance for the future of the GW field. Indeed the detection of the SGWB would not be fully successful if not accompanied by a proper  identification of its characteristics, pointing to the sources at the origin of the signal. Given the plethora of plausible SGWB sources, the better the characterization of the signal, the easier such identification will be~\footnote{The measurement of the power spectrum is not the only way to characterize the SGWB signal. See e.g.~refs.~\cite{Regimbau:2011bm,
Bartolo:2018qqn} for a discussion about the measurement of other SGWB observables.}. Our procedure is possibly one of the first steps towards the achievement of this goal.
 
\bigskip\bigskip

\noindent \textit{Note added:}
The present paper is published at the same time as Ref.~\cite{nikos}. This work proposes a methodology to assess the ability of LISA to detect a SGWB given a certain level of noise uncertainty, without the need for any simulated data. 
The main result of~\cite{nikos} are two analytic formulas, one for the probability of having excess (signal) power in a set of frequency-binned data given the LISA noise, and another on the Bayes factor of a model with signal w.r.t.~a model without signal as a function of the noise uncertainty. 
The proposed methodology is based on fixed frequency grid calculations, and Ref.~\cite{nikos} demonstrates that it works by applying it to the Radler (LISA Data Challenge) data. The approach described in the present analysis and the one of Ref.~\cite{nikos} are different but both valuable and it would be advantageous to merge them in the LISA data analysis pipeline.

\bigskip\bigskip

\noindent \textbf{Acknowledgments}\\
We thank Antoine Petiteau for useful discussions and help with the LISA noise model. We also acknowledge Robert Caldwell and Nelson Christensen  for helpful comments, and the LISA Cosmology Working Group for providing a friendly and proficient environment for the development of this work. We thank the Centro de Ciencias de Benasque Pedro Pascual, the Mainz Institute of Theoretical Physics, the University of Helsinki, and the University Autonomous of Madrid for hosting the LISA meetings where the ideas here described were discussed. 
Some of the short-term missions involved in this work were financed by the COST Action CA16104 ``Gravitational waves, black holes and fundamental physics''. 
DGF is partially supported by the ERC-AdG-2015 grant 694896 and the Swiss National Science Foundation (SNSF).
MaPi acknowledges the support of the Spanish MINECOs ``Centro de Excelencia Severo Ocho'' Programme under grant SEV-2016-0579, and has received funding from the European Unions Horizon 2020 research and innovation programme under the Marie Sk\l{}odowska-Curie grant agreement No 713366. GT is partially supported by the STFC grant ST/P00055X/1.

\bibliographystyle{JHEP}
\bibliography{bibliomanual,biblioauto}

\end{document}